\documentclass[preprint,sort&compress,12pt,3p]{elsarticle}

\usepackage{moreverb}
\usepackage{amsmath}
\usepackage{algorithmic}
\usepackage{graphics}
\usepackage{graphicx}
\usepackage{subfigure}
\usepackage{enumitem}
\usepackage{color}
\usepackage{framed}
\usepackage{wrapfig}
\usepackage{bm}
\usepackage{mathrsfs}
\usepackage{amsfonts}
\usepackage{multirow}
\usepackage{longtable}
\usepackage{hyperref}
\usepackage{paralist}
\usepackage{indentfirst}
\usepackage{relsize}
\usepackage{extarrows}
\usepackage{lineno,xcolor}
\usepackage{upgreek}
\usepackage{bm}
\usepackage{xcolor}
\usepackage{booktabs}
\usepackage[flushleft]{threeparttable}
\usepackage[linesnumbered,lined,ruled]{algorithm2e}
\usepackage{color, colortbl}
\definecolor{Gray}{gray}{0.9}

\usepackage{geometry}
\graphicspath{ {./Figure/} }

\newcommand{\eref}[1]{(\ref{#1})}

\hypersetup{
	bookmarks=true,
	bookmarksopen=true,
	bookmarksnumbered=true,
	unicode=false,
	pdftoolbar=true,
	pdfmenubar=true,
	pdffitwindow=false,
	pdfstartview={FitH},
	pdftitle={My title},
	pdfauthor={Author},
	pdfsubject={Subject},
	pdfcreator={Creator},
	pdfproducer={Producer},
	pdfkeywords={keywords},
	pdfnewwindow=true,
	colorlinks=true,
	linkcolor=blue,
	citecolor=blue,
	filecolor=magenta,
	urlcolor=blue
}

\journal{Elsevier}

\begin{document}
	
	\title{\Large Physics-guided Convolutional Neural Network (PhyCNN) for Data-driven Seismic Response Modeling}
	
	\author[NU]{Ruiyang~Zhang}
	\author[NU2]{Yang~Liu}
	\author[NU,MIT]{Hao~Sun\corref{cor}}
	\ead{h.sun@northeastern.edu}
	
	\cortext[cor]{Corresponding author. Tel: +1 617-373-3888}
	
	\address[NU]{Department of Civil and Environmental Engineering, Northeastern University, Boston, MA 02115, USA}
	\address[NU2]{Department of Mechanical and Industrial Engineering, Northeastern University, Boston, MA 02115, USA}
	\address[MIT]{Department of Civil and Environmental Engineering, MIT, Cambridge, MA 02139, USA}
	
	\begin{abstract}
		\small
	    Accurate prediction of building's response subjected to earthquakes makes possible to evaluate building performance. To this end, we leverage the recent advances in deep learning and develop a physics-guided convolutional neural network (PhyCNN) for data-driven structural seismic response modeling. The concept is to train a deep PhyCNN model based on limited seismic input-output datasets (e.g., from simulation or sensing) and physics constraints, and thus establish a surrogate model for structural response prediction. Available physics (e.g., the law of dynamics) can provide constraints to the network outputs, alleviate overfitting issues, reduce the need of big training datasets, and thus improve the robustness of the trained model for more reliable prediction. The surrogate model is then utilized for fragility analysis given certain limit state criteria. In addition, an unsupervised learning algorithm based on K-means clustering is also proposed to partition the datasets to training, validation and prediction categories, so as to maximize the use of limited datasets. The performance of PhyCNN is demonstrated through both numerical and experimental examples. Convincing results illustrate that PhyCNN is capable of accurately predicting building's seismic response in a data-driven fashion without the need of a physics-based analytical/numerical model. The PhyCNN paradigm also outperforms non-physics-guided neural networks.
	\end{abstract}
	
	
	\maketitle
	
	\section{Introduction}\label{sec1}
	Civil infrastructures are vulnerable to natural hazards such as earthquakes, tsunamis and hurricanes, especially under the effect of material aging and structural deterioration. Recent developments in advanced sensor and computation technologies provide a primary tool for monitoring structural behavior and evaluating structural integrity. The majority of existing methodologies focus on extracting structural features (e.g., modal characteristics) and updating models from the measured data, such as the Bayesian probabilistic approach \cite{yuen2010bayesian,yuen2012updating,yuen2015efficient,sun2016probabilistic,yan2017impact}, Kalman filtering \cite{yang2006adaptive,wu2007application,xie2012real}, seismic interferometry \cite{nakata2013monitoring1,nakata2013monitoring2,nakata2015damage,mordret2017continuous,sun2017bayesian}, etc. Nevertheless, how to effectively utilize sensing data for structural response modeling and prediction under future hazards remains as a challenge.
	
	Conventional approaches for structure response prediction using sensing data include identification-based (or model updating-based) methods and analytical methods. For the identification-based approach, mapping the given excitation to the corresponding response through a black-box model or state-space model \cite{sjoberg1995nonlinear,braun2002inverse,moaveni2009uncertainty,belleri2014damage,yousefianmoghadam2018system} has been used to simulate and predict the system dynamic response. A comprehensive review of works on system identification and its applications in structural response modeling was given in \cite{sohn2003review,kerschen2006past,wu2018data}. Alternatively, structural finite element (FE) model updating, which minimizes the errors between the measured response of the real structure and the synthetic response of the parametrized FE model, has been extensively studied and used to predict linear/nonlinear structural response given a new input \cite{brownjohn2000dynamic,yuen2005model,weber2009damage,song2013real,sun2015hybrid}. For instance, Skolnik et al. \cite{skolnik2006identification} predicted the seismic responses of a 15-story steel-frame building by an updated FE model. In general, these approaches require excessive computational efforts on updating the FE model when the model is of high fidelity, due to the large number of parameters for updating and limited availability of sensing data. Though low-fidelity models are more computationally cost effective, the accuracy is difficult to be retained under uncertainties especially for nonlinear response modeling. For analytical approaches, autoregressive integrated moving average (ARIMA) is one of the most popular linear models for time series analysis and forecasting \cite{fishwick1989neural,ediger2007arima}, which is originated from the autoregressive models (AR), the moving average models (MA) and the auto-regressive moving average models (ARMA). These time series modelling methods have served the scientific community for a long time; however, issues exist in regard to accuracy as well as stationarity and linearity hypothesis.
	
	Recently, considerable attention has been focused on Artificial Intelligence (AI) which has been proven to be a powerful response modeling tool and approximator \cite{irie1988capabilities,hornik1991approximation,chen1992neural,tianping1993approximations,chen1995approximation}. In particular, support vector machines (SVM) and artificial neural networks (ANN) have been used for the identification and modeling of dynamic responses during the past decade. For example, Zhang et al. \cite{zhang2007novel} employed SVM to identify structural parameters. Dong et al. \cite{yinfeng2008nonlinear} predicted the dynamic response of an oscillator and a frame structure based on a SVM-based two-stage method. In addition, multi-layer perceptron (MLP) ANN has been applied for predicting structural response under static or dynamic loading conditions. For instance, Lightbody and Irwin \cite{lightbody1996multi} applied MLP to predict the viscosity of an industrial polymerization reactor. Wang et al. \cite{ying2009artificial} developed a MLP back-propagation network to predict the seismic response of a bridge structure. Christiansen et al. \cite{christiansen2011artificial} used previous time step information as input to a one layer MLP network to predict the dynamic response of the simplified model of a wind turbine. Huang et al. \cite{huang2003neural} identified structural dynamic characteristics and performed damage diagnosis of a building using a back-propagation MLP taking previous time step information as input. Lagaros and Papadrakakis  \cite{lagaros2012neural} proposed an MLP network to predict the structural nonlinear behavior of 3D buildings when earthquake excitations with increasing intensities are considered.
	
	Nevertheless, a very limited number of studies have been reported in literature for structural response modeling and prediction using more advanced deep learning models such as the recurrent neural network (RNN) and the convolutional neural network (CNN). RNN is designed to learn sequential and time-varying patterns for regression problems \cite{mandic2001recurrent,medsker2001recurrent,yu2019aircraft,zhang2019deep}, while CNN is known for its capability in classification of data with grid-like topology (e.g., 1D sequences and 2D images) \cite{cha2017deep,atha2018evaluation,wang2018automated}. CNN can be also used for solving regression problems. Recently, Sun et al. \cite{sun2017data} proposed a virtual sensor model using CNN to estimate the dynamic responses of two numerical structure given measurements at other locations. However, the network was trained using the partial response measurements as input to predict responses of the rest of DOFs, limiting its applications in structural response prediction under new inputs. Another remarkable work by Wu and Jahanshahi \cite{wu2018deep} used CNN to estimate structural dynamic response and perform system identification, which showed great capability of CNN for sequence regression. Nevertheless, the hypothesis was that the data is sufficient to train a reliable predictive model. Challenges arise when the available training data is scarce.
	
	We herein address limitations in data-driven structural response modeling through developing a novel physics-guided CNN (i.e., PhyCNN), which is capable of accurately predicting nonlinear structural seismic time-history responses in a data-driven manner. The basic concept is to (1) embed available physics knowledge into the deep learning model, (2) train a PhyCNN based on available seismic input-output datasets (e.g., from simulation or sensing), and (3) use the trained PhyCNN as a surrogate model for response prediction. The surrogate model can further be utilized for fragility analysis given certain limit state criteria (e.g., the serviceability state). It is noted that the available physics (e.g., the law off dynamics) can provide constraints to the network outputs, alleviate overfitting issues, reduce the need of big training datasets, and thus improve the robustness of the trained model for more reliable prediction.
	
	This paper is organized as follows. Section \ref{sec2} presents the proposed PhyCNN architecture for structural response modeling. In Section \ref{sec3}, the performance of PhyCNN is verified through two numerical examples of a nonlinear system. Section \ref{sec4} presents the experimental validation of PhyCNN based on filed sensing measurements, where data-driven serviceability analysis of a building is also discussed. Section \ref{sec_con} summarizes the conclusions.
	
	\section{Physics-guided Convolutional Neural Network (PhyCNN)}\label{sec2}
	Neural networks have been widely recognized as a powerful tool to deal with problems like classification and regression. Among many other neural networks, CNN, which is inspired by the virtual convex of animals \cite{ciresan2011flexible}, can effectively model the grid-structured topology of data (e.g., images), making it especially powerful for image classification. However, CNN is also capable of dealing with regression problems, which is often inconspicuous due to its capability in classification. Traditionally, deep neural networks are trained solely based on data. However, by adding physics (e.g., the governing law of dynamics) into the training phase, the robustness and reliability of learning from the data can be further enhanced. In other words, the embedded physics can inform the learning and constrain the training to a feasible space. In this paper, a 1D regression-oriented PhyCNN architecture is proposed for time series modeling.
	
	To illustrate the concept, let's consider a dynamic system subjected to the ground excitation following the equation of motion below:
	\begin{equation}
	\label{eq:EOM}
	\textbf{M}\ddot{\mathbf{x}}(t) + \mathbf{h}(t) = -\textbf{M}\Gamma \ddot{\mathbf{x}}_\mathrm{g}(t)
	\vspace{0mm}
	\end{equation}
	where \textbf{M} is the mass matrices; \(\mathbf{x}\), \(\dot{\mathbf{x}}\), and \(\ddot{\mathbf{x}}\) are the relative displacement, velocity, and acceleration vectors to the ground; \(\ddot{\mathbf{x}}_\mathrm{g}\) represents the ground acceleration; \(\Gamma\) is the force distribution vector; and \(\mathbf{h}\) is the generalized restoring force vector. Normalizing Eq. \eref{eq:EOM} by \textbf{M}, the governing equation can be expressed as
	\begin{equation}
	\label{eq:eom}
	f:=\ddot{\mathbf{x}}(t) + \mathbf{g}(t) + \Gamma \ddot{\mathbf{x}}_\mathrm{g}(t) \xrightarrow[]{}0
	\vspace{0mm}
	\end{equation}
	where \(\mathbf{g}(t)\) is the mass-normalized restoring force, namely, $\mathbf{g}(t) = \textbf{M}^{-1}\mathbf{h}(t)$.
	
	\begin{figure}[t!]
		\centering\includegraphics[width=1\linewidth]{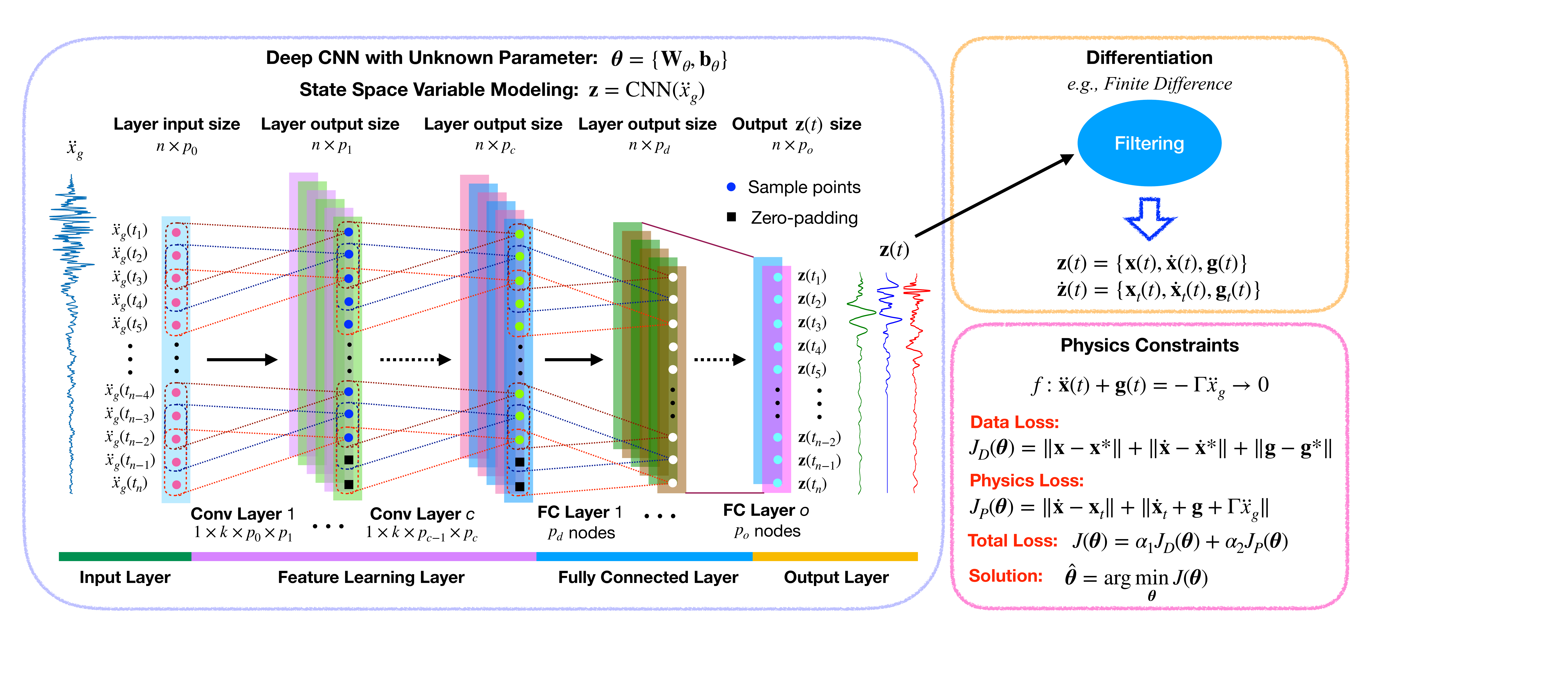}
		\caption{The proposed physics-guided convolutional neural network (PhyCNN) for time-series modeling. The PhyCNN architecture includes the input layer, the feature learning layer, fully-connected layer, the output layer, and the graph-based tensor differentiator. The inputs are ground accelerations (or ground displacements) and the outputs are state space variables $\mathbf{z}(t)$ including displacement $\mathbf{x}(t)$, velocity $\dot{\mathbf{x}}(t)$, and restoring force $\mathbf{g}(t)$, namely, $\mathbf{z}(t)=\{\mathbf{x}_t(t), \dot{\mathbf{x}}_t(t), \mathbf{g}_t(t)\}$. The derivatives of state space outputs $\mathbf{\dot{z}}(t)$ are calculated through a tensor differentiator using the finite difference method. The total loss consists of the data loss from the measurements and the physics loss which models the dependency between the output features. Both input and output to the network are time sequences, including $p_0$ (e.g., ground acceleration and/or ground displacement) and $p_o$ (e.g., state space variables at different floor levels) features, respectively. The size of layer input and output is given. The convolution layer is defined as ``height $\times$ width $\times$ depth $\times$ filters (output channels)''. An identical kernel size is used for all three convolution layers in this study. Zero-padding is added to the output sequence of each convolution layer due to the convolution operation as illustrated in Section \ref{sec2.1}. Note that the nonlinear activation functions are not shown in this figure.}
		\label{fig:cnn}
	\end{figure}
	
	A PhyCNN framework is developed for surrogate modeling of such a nonlinear dynamic system under ground motion excitation. The proposed deep learning framework consists of a 1D CNN and a graph-based tensor differentiator. Figure \ref{fig:cnn} shows the basic concept and architecture of PhyCNN in the context of structural response modeling given the ground acceleration as input which contains n sample points from $t_1$ to $t_n$. The outputs are state space variables $\mathbf{z}(t)$ including the structural displacement $\mathbf{x}(t)$, the velocity $\dot{\mathbf{x}}(t)$, and the normalized restoring force $\mathbf{g}(t)$, namely, $\mathbf{z}(t)=\{\mathbf{x}(t), \dot{\mathbf{x}}_t(t), \mathbf{g}(t)\}$, each of which has same number of $n$ sample points ranging from $t_1$ to $t_n$. With the convolution operation illustrated in Section \ref{sec2.1}, zero-padding is added to the output sequence of each convolution layer to ensure the identical input/output length. The proposed PhyCNN architecture consists of multiple hidden layers besides the input and output layers, namely, the feature learning layers and the fully connected layers \cite{lawrence1997face,lecun2015deep}. A typical feature leaning layer usually includes a convolution layer, a nonlinear layer (or nonlinear activation function), and a feature pooling layer. The output of each layer is called a feature map since the feature learning layers are used for extracting features from the input or from the output from the previous layer. In the proposed PhyCNN architecture, the dimension of height in a classical CNN is reduced to one, making it possible to take a time-series signal as input, while the dimension of width represents the temporal space. In addition, a graph-based tensor differentiator (e.g., the finite different method) is developed to calculate the derivative of state space outputs $\dot{\mathbf{z}}(t)=\{\mathbf{x}_t(t), \dot{\mathbf{x}}_t(t), \mathbf{g}_t(t)\}$ to construct the physics loss from the governing equation, where the subscript $t$ represents the derivative of the state with respect to time. The basic concept here is to optimize the network hyperparameters $\boldsymbol{\theta} = \{\mathbf{W}_\theta, \mathbf{b}_\theta\}$ such that the PhyCNN can interpret the measurement data (e.g., $\mathbf{x}^m,\dot{\mathbf{x}}^m,\mathbf{g}^m$) while satisfying the physical equation of motion in Eq. \eref{eq:eom}, e.g., $f\xrightarrow[]{}0$. Here, $\mathbf{W}_\theta$ and $\mathbf{b}_\theta$ are the neural network weight and bias parameters. The total loss $J\left(\mathbf{\boldsymbol{\theta}}\right)$ is then defined as
	\begin{equation}\label{eq:loss}
	J\left(\mathbf{\boldsymbol{\theta}}\right)=J_D\left(\mathbf{\boldsymbol{\theta}}\right) + J_P\left(\mathbf{\boldsymbol{\theta}}\right)
	\vspace{1pt}
	\vspace{0mm}
	\end{equation}
	with
	\begin{equation}\label{eq:loss_D}
	J_D\left(\mathbf{\boldsymbol{\theta}}\right)=\frac{1}{N}\left\Vert\mathbf{x}^{p}-\mathbf{x}^{m}\right\Vert_{2}^{2}+\frac{1}{N}\left\Vert\dot{\mathbf{x}}^{p}-\dot{\mathbf{x}}^{m}\right\Vert_{2}^{2}+\frac{1}{N}\left\Vert\mathbf{g}^{p}-\mathbf{g}^{m}\right\Vert_{2}^{2}
	\vspace{1pt}
	\end{equation}
	\vspace{-3mm}
	\begin{equation}\label{eq:loss_P}
	J_P\left(\mathbf{\boldsymbol{\theta}}\right)=\frac{1}{N}\left\Vert\dot{\mathbf{x}}^{p}-\mathbf{x}_t^{p}\right\Vert_{2}^{2}+\frac{1}{N}\left\Vert\dot{\mathbf{x}}_t^{p}+\mathbf{g}^{p}+\Gamma\ddot{\mathbf{x}}_\mathrm{g}\right\Vert_{2}^{2}
	\vspace{1pt}
	\end{equation}
	where $J_D\left(\mathbf{\boldsymbol{\theta}}\right)$ denotes the data loss based the measurements while $J_P\left(\mathbf{\boldsymbol{\theta}}\right)$ represents the physics loss which introduces a constraint for the neural network that models the dependency in-between the output features; the superscript $p$ and $m$ denote the prediction and measurement, respectively. Note that the measurements are not necessarily required for the complete state, which could be part of the state variables (e.g., $\mathbf{x}^{m}$ only) or the accelerations (e.g., $\ddot{\mathbf{x}}^{m})$. In such a case, the data loss in Eq. \eref{eq:loss_D} should be adjusted accordingly. During training, $\mathbf{\boldsymbol{\theta}}$ will be updated and determined by solving the optimization problem $\mathbf{\hat{\boldsymbol{\theta}}}:= \operatorname*{arg\, min}_{\mathbf{\boldsymbol{\theta}}} J\left(\mathbf{\boldsymbol{\theta}}\right)$. Note that the proposed PhyCNN architecture used in this study has five convolution layers ($c=5$) with identical kernel size and three fully-connected layers. Details of each layer are discussed in the following subsections.
	
	\subsection{Convolution Layer}\label{sec2.1}
	The convolution (Conv) layer is the basis of the CNN architecture, which performs the core operations of feature learning. The size of a convolution layer is defined as ``height $\times$ width $\times$ depth $\times$ filters (output channels)'', e.g., $1\times k \times p_0\times p_1$ for the first Conv layer as shown in Figure \ref{fig:cnn}. Each Conv layer consists of a set of learnable kernels (also known as filters) with a size of $1\times k$, which are parameterized by a hyperparameter called the receptive field containing a group of weights shared over the entire input temporal space. The initial weights of a receptive field are typically randomly generated. During the forward pass, the kernels convolve across the temporal space of the input and compute dot products between the entries of a receptive field and a local region of the input which represents a sequence of input time series. The dot products are summed, and the bias is added to the summed value, forming a single entry $z_{ij}^{\left(l\right)}$ of the output. The full input space is scanned through sliding the kernels along the temporal space with one single stride. In this way, the time dependency is captured by convolving a sequence of input across the entire temporal space. The stride operation will lead to smaller output length if zero padding is not applied. To ensure the output has the same length $n$ as the input in the temporal space, zero-padding is added at the end of the input time series. The number of required zero-padding, $P$, is given by $P=k-1$. The dimension of the convolution layer output $z_i^{\left(l\right)}$ can be different from the layer input $z_i^{\left(l-1\right)}$, where $l$ denotes the layer index. The number of filters, $p$, specifies the dimensionality of the output space. For an input feature $j$ (i.e., or called the channels in a standard CNN), the corresponding output of a convolution layer can be written as 
	\begin{equation}
	\label{eq:cnn}
	z_{ij}^{(l)} = \sum_{i}^{i+k-1} W_{j}^{(l)} \ast z_{i}^{(l-1)} + b_{ij}^{(l)}
	\vspace{0mm}
	\end{equation}
	which takes the output of the previous layer $z_i^{\left(l-1\right)}$ with zero-padding as input. Here, $i$ represents the time step in the temporal space $(i=1,\ 2,\ \cdots,n)$; $W_j^{\left(l\right)}$ is the receptive field with the kernel size of $k$; $b_{ij}^{\left(l\right)}$ is the bias added to the summed term; and $\ast$ represents the 1D convolution operator. 
	
\begin{figure}[b!]
		\centering\includegraphics[width=0.85\linewidth]{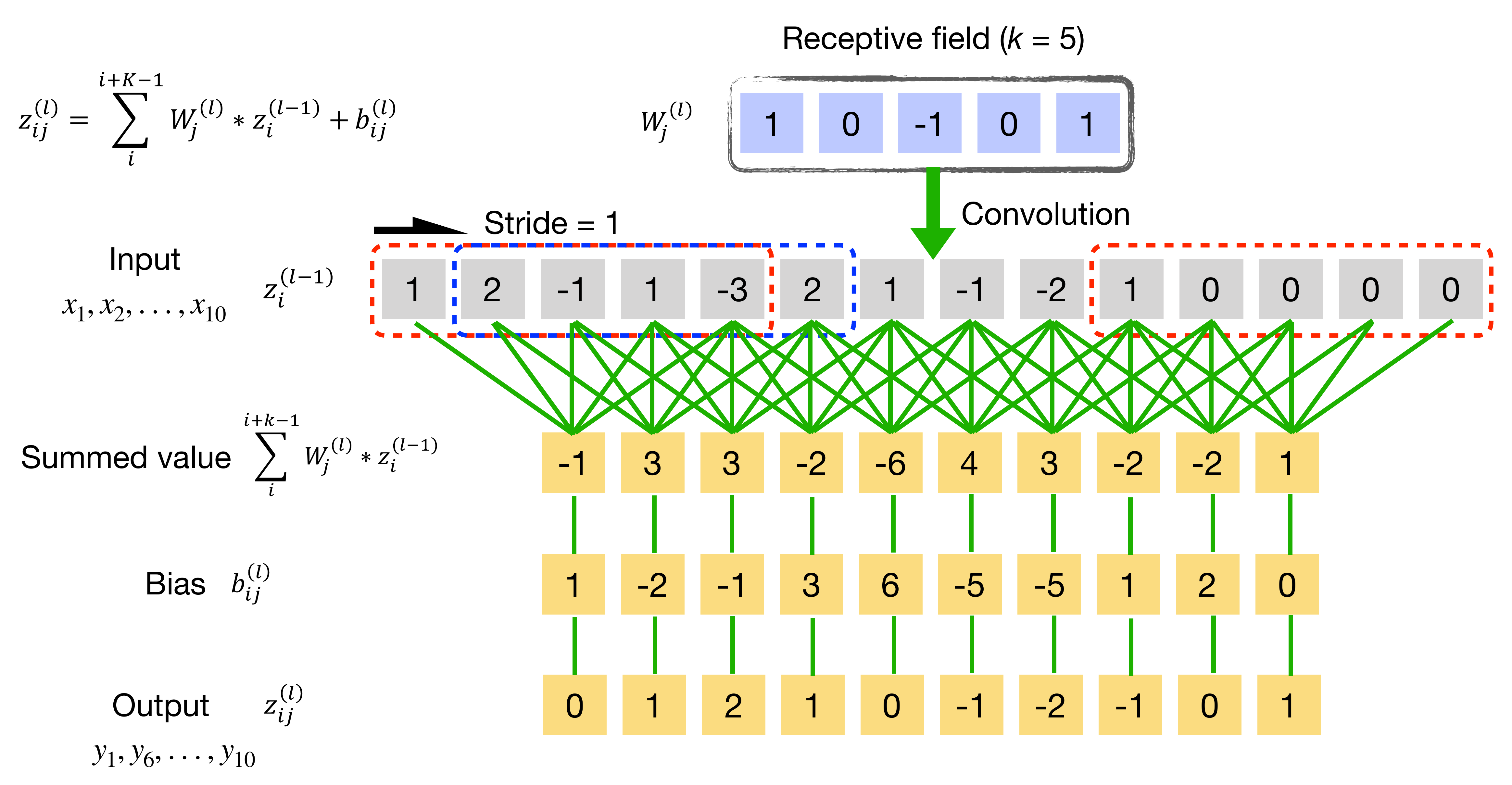}
		\caption{Illustration of the convolution process in a single convolution layer with $n$ = 10, $k$ = 5 and $s$ = 1.}
		\label{fig:conv}
	\end{figure}

	A simple example is presented in Figure \ref{fig:conv} to illustrate the convolution operation in the temporal space. Herein, an input sequence with a length of $n$ = 10 is considered, with only one filter and a receptive filed size of $k$ = 5. The kernel slides across the temporal space with a stride $s$ = 1, resulting in the output with a length of 10. The weights of the receptive field are [1, 0, \(-1\), 0, 1], shared across the entire temporal space.
	
	Conventionally, a convolution layer is always followed by a nonlinear activation function, which introduces nonlinearity and makes learning easier by adapting with variety of data and differentiating between the output. The activation layer increases the nonlinearities of the model and the overall network without affecting the receptive fields of the convolution layer. Common nonlinear activation functions include rectified linear unit (ReLU) \cite{nair2010rectified}, hyperbolic tangent (Tanh), and sigmoid function. In this paper, the ReLU function shown in Eq. \eref{eq:relu} is employed, which gives slightly better performance compared to Tanh and sigmoid.
	
	\begin{equation}
	\label{eq:relu}
	f(x) =
	\begin{cases}
	0, ~~\mathrm{for}~~ x < 0\\
	x, ~~\mathrm{for}~~ x \geq 0
	\end{cases}
	\end{equation}
	\vspace{0mm}
	
	\subsection{Pooling Layer}
\begin{figure}[b!]
		\centering\includegraphics[width=0.75\linewidth]{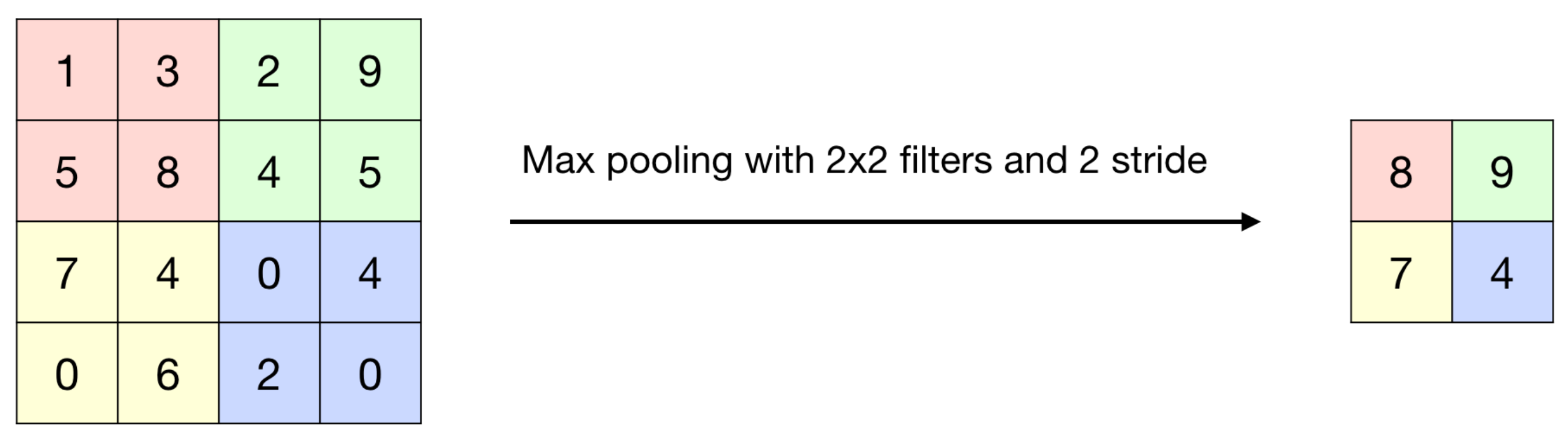}
		\caption{An example of max pooling operation.}
		\label{fig:pooling}
	\end{figure}

	The pooling layer is often used to reduce the spatial size of the feature maps when dealing with classification problems with large input data. Popular pooling layers include max pooling and average pooling, which take either maximum or mean values from a pooling window. For example, Figure \ref{fig:pooling} shows the max pooling operation in a standard CNN. In PhyCNN, pooling layer will perform a down-sampling operation in the temporal space, resulting in smaller output length, which is undesired for regression problems like time-series prediction. Although zero-padding can be applied to keep same output length as input, the time dependencies are alternated. Therefore, the pooling layer is restricted in the proposed PhyCNN architecture for structural response modeling.
	
	\subsection{Fully-Connected Layer}
	Exactly as its name implies, the fully-connected (FC) layer has full connections to all activations in the previous layer, as observed in regular neural networks. A FC layer multiplies the input by a weight matrix and then adds a bias vector. FC layers are typically used in the last stage of the CNN to connect to the target output layer and construct the desired number of output classes.
	For regression problems like time-series prediction, nonlinear activation functions such as ReLU, Tanh, and sigmoid are inappropriate for the last FC layer since they map the output in the range of [0, infinite), ($-$1, 1), and (0, 1), respectively. Other potential alternatives include parametric rectified linear unit (PReLU) \cite{xu2015empirical} and exponential linear unit (ELU) \cite{trottier2017parametric}. In this paper, the Tanh function is used as the activation within the FC layers and the linear activation function is applied for the output layer.
	
	\subsection{Dropout Layer}
	Dropout layers can be added after each convolution layer and fully-connected layer to reduce overfitting by preventing complex co-adaptations on training data \cite{srivastava2014dropout}, which has remained as a common issue in machine learning. The key idea is to randomly disconnect the connections and drop units from the connected layer with a certain dropout rate during training. Dropout layers can also improve the training speed. Typically, they are applied before the FC layer which has more learnable parameters and is more likely to cause overfitting. Although convolution layers are less likely to overfit due to their particular structure where weights are shared over the spatial space, dropout layers can still be applied to the convolution layers which have huge parameters. In this study, dropout layers with a dropout rate of 0.2 are applied before the FC layers.
	
	\subsection{PhyCNN for Response Modeling}
	The proposed PhyCNN architecture takes the ground motion (e.g., ground accelerations) as input and the structural responses (e.g., story displacements) as output to learn the feature mapping between the input and output. First, the model is trained with either synthetic database or field sensing measurements. Then the trained model can be used to predict the structural responses under new seismic excitations. To train the proposed PhyCNN architecture, both the input and output dataset must be formatted as a three-dimensional array, where the entries are samples in the first dimension, time history steps in the second dimension, and input or output features in the last dimension. The detailed neural network architecture is illustrated in Figure \ref{fig:cnn}, including five convolution layers and three fully-connected layers in addition to the input and output layer. Each convolution layer has 64 filters with a kernel size of 50 used in this study. The number of filters (nodes) and kernel length can be adjusted to get better performance for different problems. Note that the number of nodes for the last FC layer must be equal to the number of output features.
	
	The entire training process is performed in a Python environment using Keras \cite{chollet2015keras}. Keras is a high-level open source deep learning library built on top of TensorFlow which offers easy and fast prototyping neural networks. TensorFlow, a symbolic math library for machine learning applications developed by Google Brain Team \cite{abadi2016tensorflow}, is served as the backend engine in Keras. It offers flexible data flow architecture enabling high-performance training of various types of neural networks across a variety of platforms (CPUs, GPUs, TPUs). The simulations are performed on a standard PC with 28 Intel Core i9-7940X CPUs and 2 NVIDIA GTX 1080 Ti video cards. The data and codes used in this paper will be publicly available on GitHub at \href{https://github.com/zhry10/PhyCNN}{https://github.com/zhry10/PhyCNN} after the paper is published.
	
	\section{Numerical Validation}\label{sec3}
	We first demonstrate the performance of the PhyCNN approach for predicting structural displacements through two numerical examples. In the first example, we assume that the field measurements of all states, including $\mathbf{x}$, $\dot{\mathbf{x}}$, and $\mathbf{g}$, are available for training. Note that $\mathbf{g}$ can be inferred from the measurement of $\ddot{\mathbf{x}}$, e.g., $\mathbf{g} = -\ddot{\mathbf{x}}-\Gamma\ddot{\mathbf{x}}_\mathrm{g}$. Noteworthy, these responses can be recorded from numerical simulations when using PhyCNN for reduced order surrogate modeling. In the second example, only the measurements of $\ddot{\mathbf{x}}$ are available for training. The aim of having the aforementioned two scenarios is to show the versatility of the proposed PhyCNN for dealing with various availability of measurement data, which, for example, takes account for the access of field sensing measurements in real applications. In both examples, a single degree-of-freedom (DOF) nonlinear system subjected to ground motion excitation is investigated, whose equation of motion of the nonlinear system is expressed as:
	\begin{equation}
	m\ddot{x}+\underbrace{c\dot{x}+k_1x+k_2x^3}_{h} = -m\mathbf{\Gamma}{\ddot{x}}_\mathrm{g}
	\end{equation}
	where $m=1$ kg is the mass, $c=1$ Ns/m is the damping coefficient, $k_1=20$ N/m is the linear stiffness coefficient, and $k_2=200$ N/m is the nonlinear stiffness coefficient. The mass-normalized restoring force reads $g=h/m$.
	
	\subsection{Case 1: Available Measurements of $\mathbf{x}$, $\dot{\mathbf{x}}$, $\mathbf{g}$}\label{sec3.1}
	A synthetic database, consisting of 100 samples (i.e., independent seismic sequences), was generated by numerical simulation of the 1DOF nonlinear system excited by a suite of earthquake records selected from the PEER strong motion database \cite{chiou2008nga} with a 10\% probability of exceedance in 50 years. Each simulation was executed up to 50 seconds with a sampling frequency of 20 Hz resulting in 1,001 data points for each record. However, only 10 datasets are randomly selected and considered as known datasets for training, while the rest are considered as unknown datasets to show the prediction performance. The PhyCNN architecture shown in Figure \ref{fig:cnn} is implemented to develop the surrogate model. The input and output are formatted into the shape of [10, 1001, 1] and [10, 1001, 3] for the training datasets. To show the performance of the proposed approach with physics constraint, our method is also compared with the regular CNN without the physics loss (denoted as CNN). 
	
\begin{figure}[t!]
		\centering\includegraphics[width=0.975\linewidth]{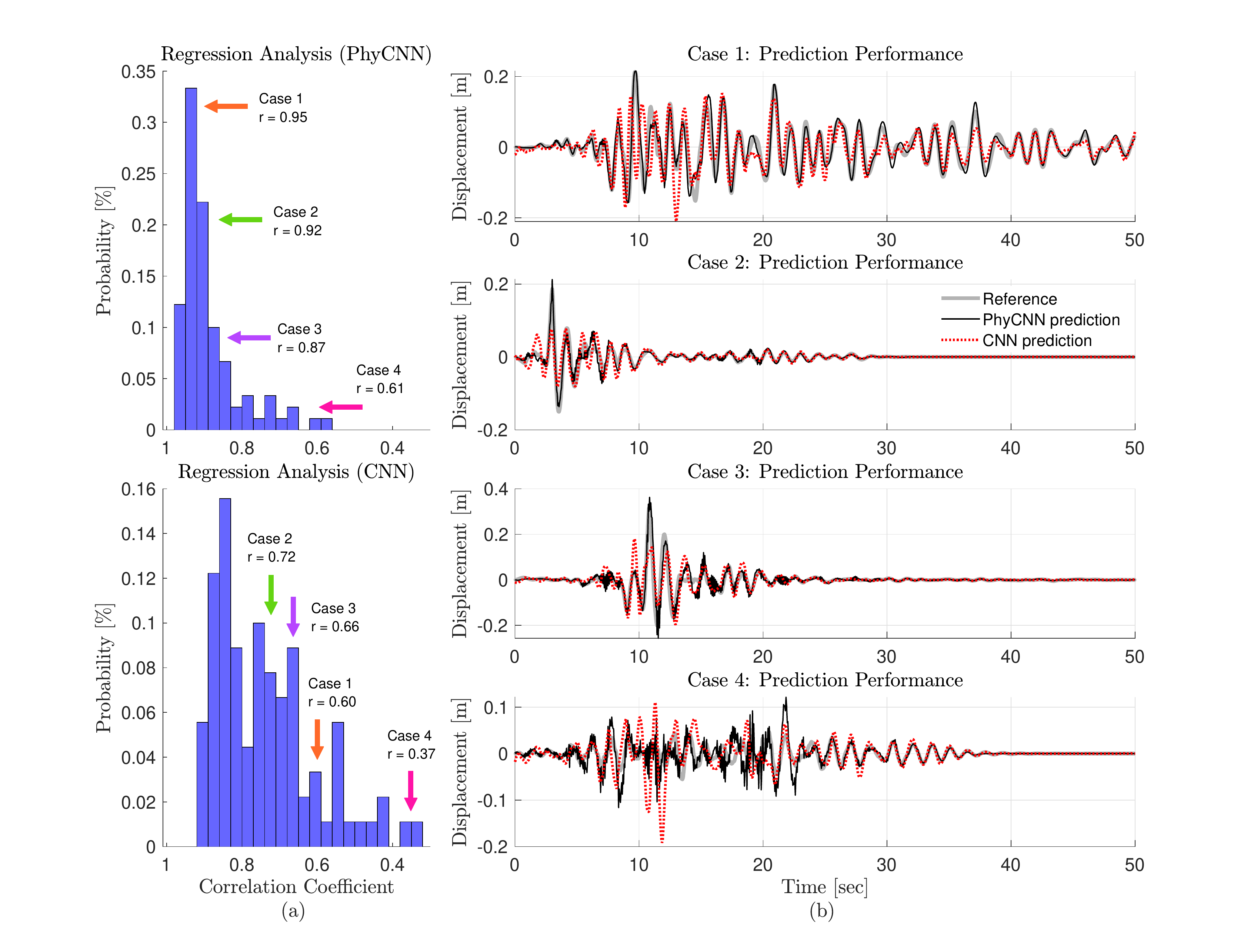}
		\vspace{-16pt}
		\caption{Regression analysis in (a) and four examples of predicted displacements in (b) for unknown earthquakes using PhyCNN and CNN.}
		\label{fig:reg_u}
	\end{figure}

\begin{figure}[t!]
		\centering\includegraphics[width=1\linewidth]{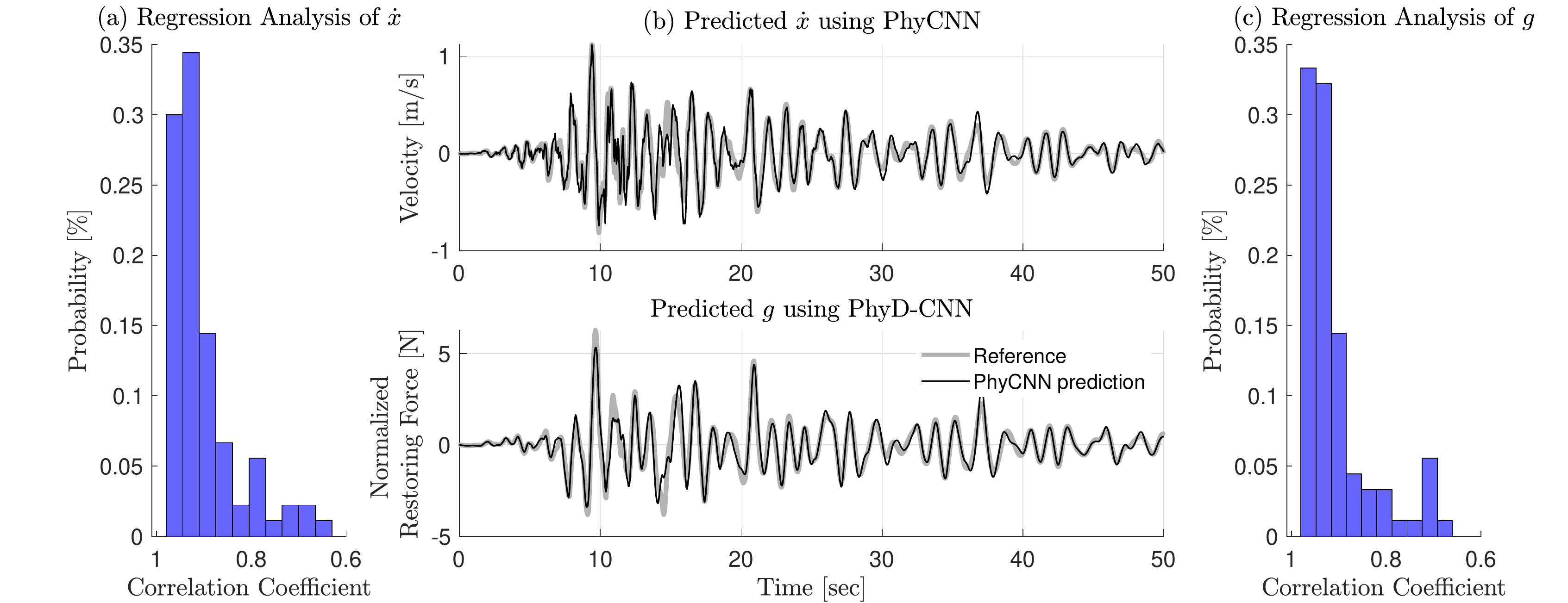}
		\caption{Prediction performance of $\dot{x}$ and $g$ using PhyCNN: (a, c) regression analysis of $\dot{x}$ and $g$ respectively; (b) an example of predicted time history of $\dot{x}$ and $g$.}
		\label{fig:reg_ut_g}
	\end{figure}

\begin{figure}[t!]
		\centering\includegraphics[width=0.75\linewidth]{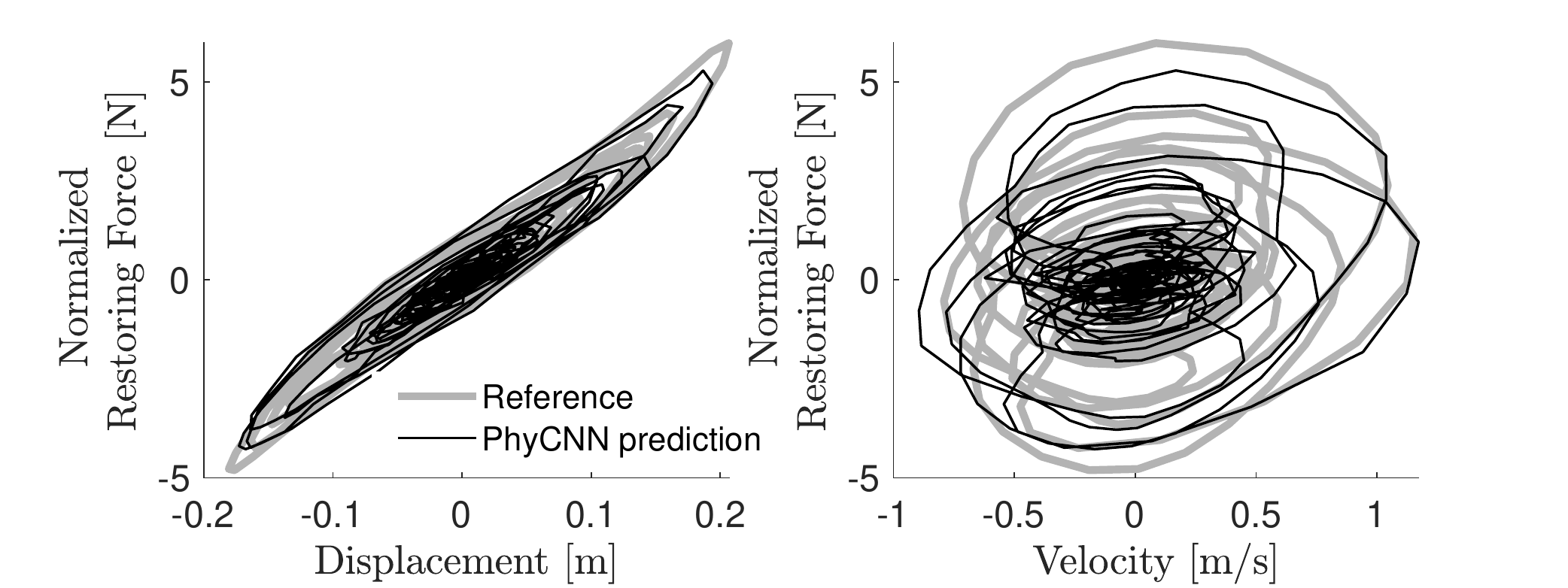}
		\caption{Predicted hysteresis of nonlinear restoring force versus displacement (left) and nonlinear restoring force versus velocity (right).}
		\label{fig:hystere}
	\end{figure}

	Figure \ref{fig:reg_u} summarizes the prediction performance of both PhyCNN and CNN. Figure \ref{fig:reg_u}(a) shows the regression analysis across all 90 prediction datasets for both PhyCNN (top-left) and CNN (bottom-left). It can be clearly seen that the prediction accuracy is greatly increased by embedding the physics constraints into deep learning. The time histories of predicted displacements are presented in Figure \ref{fig:reg_u}(b) corresponding four different levels of correlation coefficients (noted by $r$), namely, 0.95, 0.92, 0.87, 0.61 using PhyCNN and 0.60, 0.72, 0.66, 0.37 using CNN. The PhyCNN prediction matches the reference well in both magnitudes and phases. Even for the worst case of $r=0.61$, the proposed PhyCNN approach is able to reasonably predict the structural dynamics. On the contrary, CNN produces less satisfactory prediction especially in predicting the displacement magnitudes. Another salient feature of PhyCNN is that it also accurately predicts the states of velocity $\dot{x}$ and nonlinear restoring force $g$, as illustrated by the regression analysis in Figure \ref{fig:reg_ut_g}(a) and (c). Figure \ref{fig:reg_ut_g} (b) shows examples of predicted time histories of $\dot{x}$ and $g$ using PhyCNN indicating a good agreement with the ground truth. The predicted nonlinearity is given in Figure \ref{fig:hystere} which shows an example of the predicted hysteresis of the normalized nonlinear restoring force versus displacement and velocity.

	\subsection{Case 2: Available Measurements of $\ddot{\mathbf{x}}$ only}\label{sec3.2}
	In most engineering practices, only accelerometers are installed on the building to record acceleration time histories. In such cases, it is impossible to train a standard deep learning model to predict structural displacements using acceleration measurements only. One way is to calculate the displacements from the acceleration measurements first and then train a deep learning model based on the interpreted displacements. However, it is well known that inevitable large numerical errors exist due to the integration of accelerations to displacements. This somehow limits the application of standard neural networks such as CNN in real-world applications.
	
	\begin{figure}[t!]
		\centering\includegraphics[width=1\linewidth]{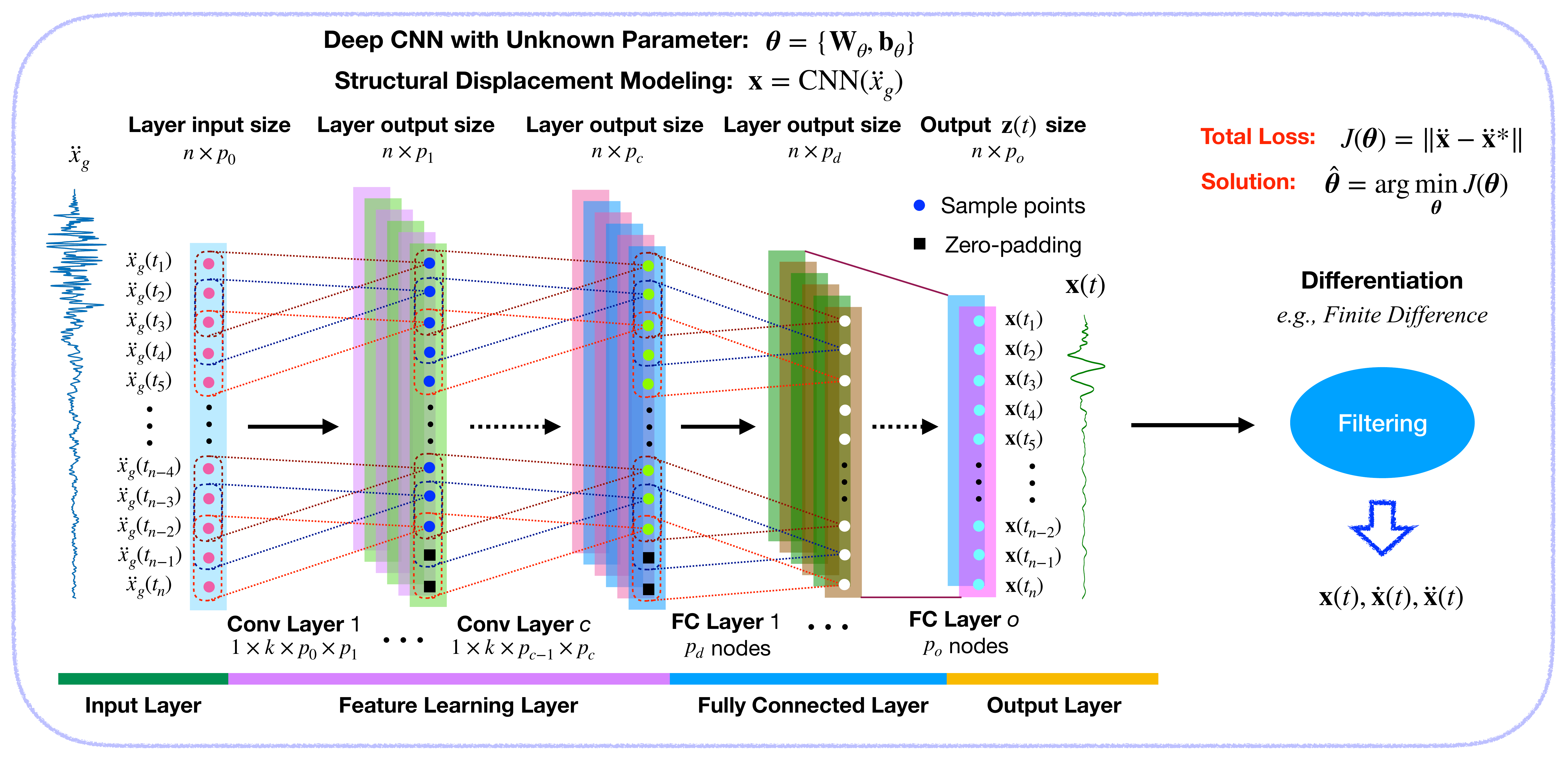}
		\caption{The modified PhyCNN for structural displacement prediction without displacement measurements for training. The only available measurements are the structural accelerations which are used to train the PhyCNN model.}
		\label{fig:cnn_ag2utt}
	\end{figure}
	
	However, with physics embedded into the training, the proposed PhyCNN is capable of accurately predicting the displacements based on only the acceleration measurements for training. This is realized by the graph-based tensor differentiator to construct the physics graph. Figure \ref{fig:cnn_ag2utt} shows the network architecture specified in this study which is similar to the general state space format used in the previous study as shown in Figure \ref{fig:cnn}. The input to the CNN is the ground motion and the output displacement is passed into the differentiator to calculate the acceleration. The model will be trained and optimized to minimize the objective function defined as follows based on acceleration measurements. 
	\begin{equation}\label{eq:loss_ag2utt}
	J\left(\mathbf{\boldsymbol{\theta}}\right)=\frac{1}{N}\left\Vert\ddot{\mathbf{x}}^{p}-\ddot{\mathbf{x}}^{m}\right\Vert_{2}^{2}
	\end{equation}
	
	\begin{figure}[t!]
		\centering\includegraphics[width=1\linewidth]{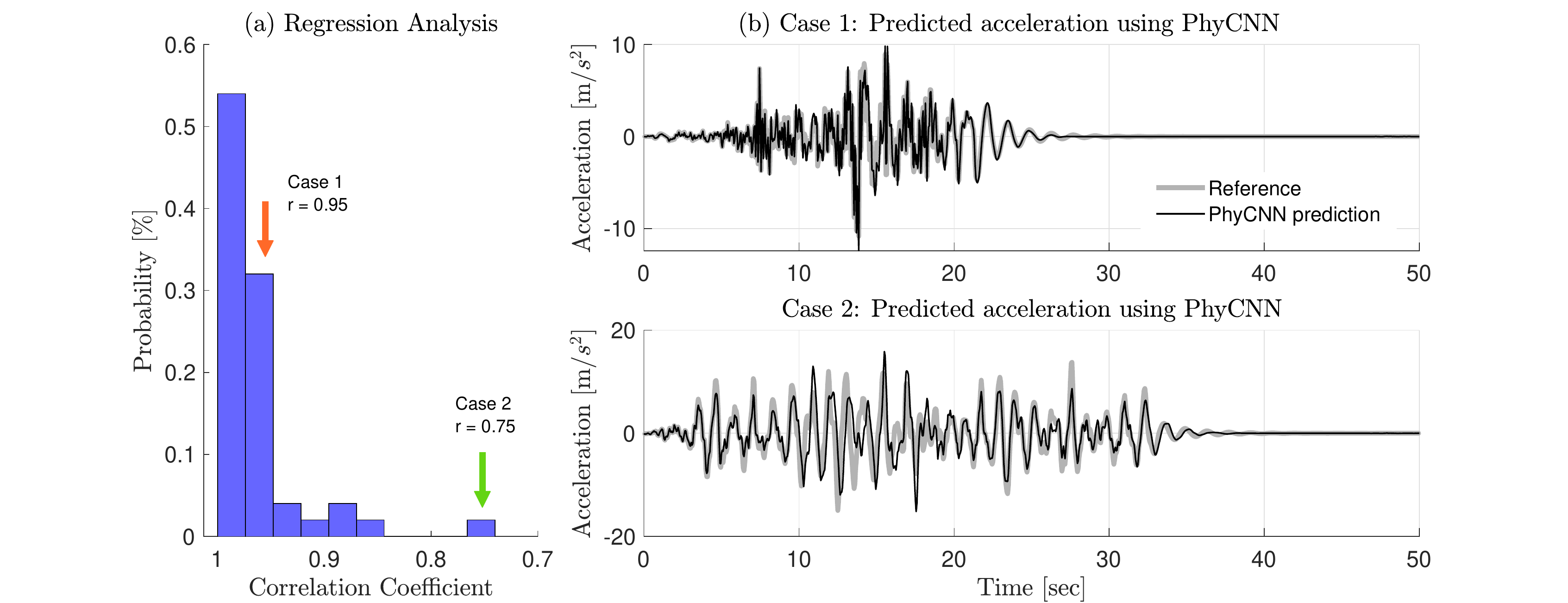}
		\caption{Prediction performance of structural acceleration $\ddot{x}$ using PhyCNN: (a) regression analysis; (b) examples of predicted time history of $\ddot{x}$.}
		\label{fig:reg_ag2utt_utt}
	\end{figure}
	
\begin{figure}[t!]
		\centering\includegraphics[width=1\linewidth]{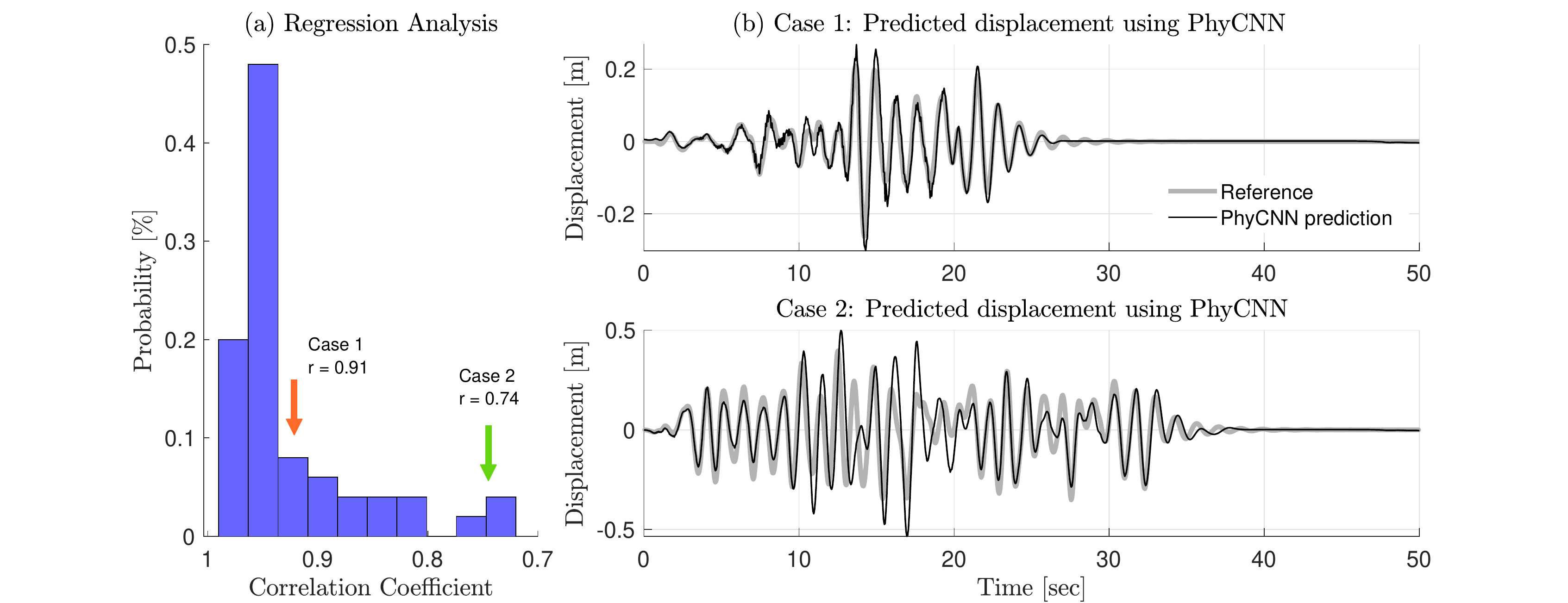}
		\caption{Prediction performance of structural displacement $x$ using PhyCNN: (a) regression analysis; (b) examples of predicted time history of $x$.}
		\label{fig:reg_ag2utt_u}
	\end{figure}

	In this case, only acceleration measurements are considered as known and used to calculate the loss. The model is trained using 50 randomly selected datasets and tested with the rest 50 datasets considered as unknown. Figure \ref{fig:reg_ag2utt_utt} shows the predicted acceleration compared to the ground truth. It is seen that the majority of correlation coefficients are greater than 0.9. Two example acceleration time histories are presented in Figure \ref{fig:reg_ag2utt_utt}(b) with different levels of accuracy. Even for the worst case with $r=0.75$, a reasonably good matching in both magnitudes and phases is observed between the PhyCNN prediction and the ground truth. Figure \ref{fig:reg_ag2utt_u} shows the predicted displacements of the nonlinear system. From the regression analysis in Figure \ref{fig:reg_ag2utt_u}(a), the correlation coefficients are mainly greater than 0.8. Figure \ref{fig:reg_ag2utt_u}(b) shows the comparison of displacement time histories. It can be clearly seen that the PhyCNN can produce accurate displacement prediction even under the circumstances that only limited acceleration measurements are available for training. This study clearly demonstrates another benefit of the proposed PhyCNN in structural response modeling in the context of latent response prediction.

	\section{Experimental Validation of PhyCNN Performance}\label{sec4}
	The PhyCNN architecture is further demonstrated using filed sensing data. A 6-story hotel building in San Bernardino, CA, from the Center for Engineering Strong Motion Data (CESMD), is selected and investigated \cite{haddadi2008center}. The deep PhyCNN model illustrated in Figure \ref{fig:cnn} was trained for the instrumented building with ground accelerations as input and the structural displacements as output. However, the measurement data used to train the PhyCNN are the acceleration time histories only (referred to the modified PhyCNN in Figure \ref{fig:cnn_ag2utt}). A data clustering technique is proposed to partition the data sets for training, validation and prediction. Based on the trained PhyCNN model, the serviceability of the 6-story hotel building can be further analyzed given new seismic inputs.
	
	\subsection{6-Story Hotel Building in San Bernardino, CA}
	The 6-story hotel building in San Bernardino, California (CA) is a mid-rise concrete building designed in 1970 with a total of nine accelerometers installed on the 1st floor, 3rd and roof floors in both directions. The sensors, with their locations shown in Figure 6, have recorded multiple seismic events in history from 1987 to 2018. Table \ref{table1} summarizes a total of 23 available datasets on CESMD used in this example. The historically recorded data is then used to train the PhyCNN. The trained surrogate model is then used to predict structural displacement time histories given new ground motions and develop a fragility function for serviceability assessment of the building.
	
	Selecting training/validation datasets plays a critical role in deep learning. Commonly, the database is divided into training, validation and prediction datasets randomly (e.g., with ratios of 70\%, 15\%, 15\%, respectively). In the case when the database is very limited, dataset partition could have a significant influence on the generalizability of the trained model. To extensively utilize the limited sensing data in this study, an unsupervised learning technique based on the K-means algorithm is proposed for data clustering. To better illustrate the concept, the 6-story hotel building in San Bernardino, CA is taken as an example.
	
	\begin{figure}[t!]
		\centering\includegraphics[width=0.75\linewidth]{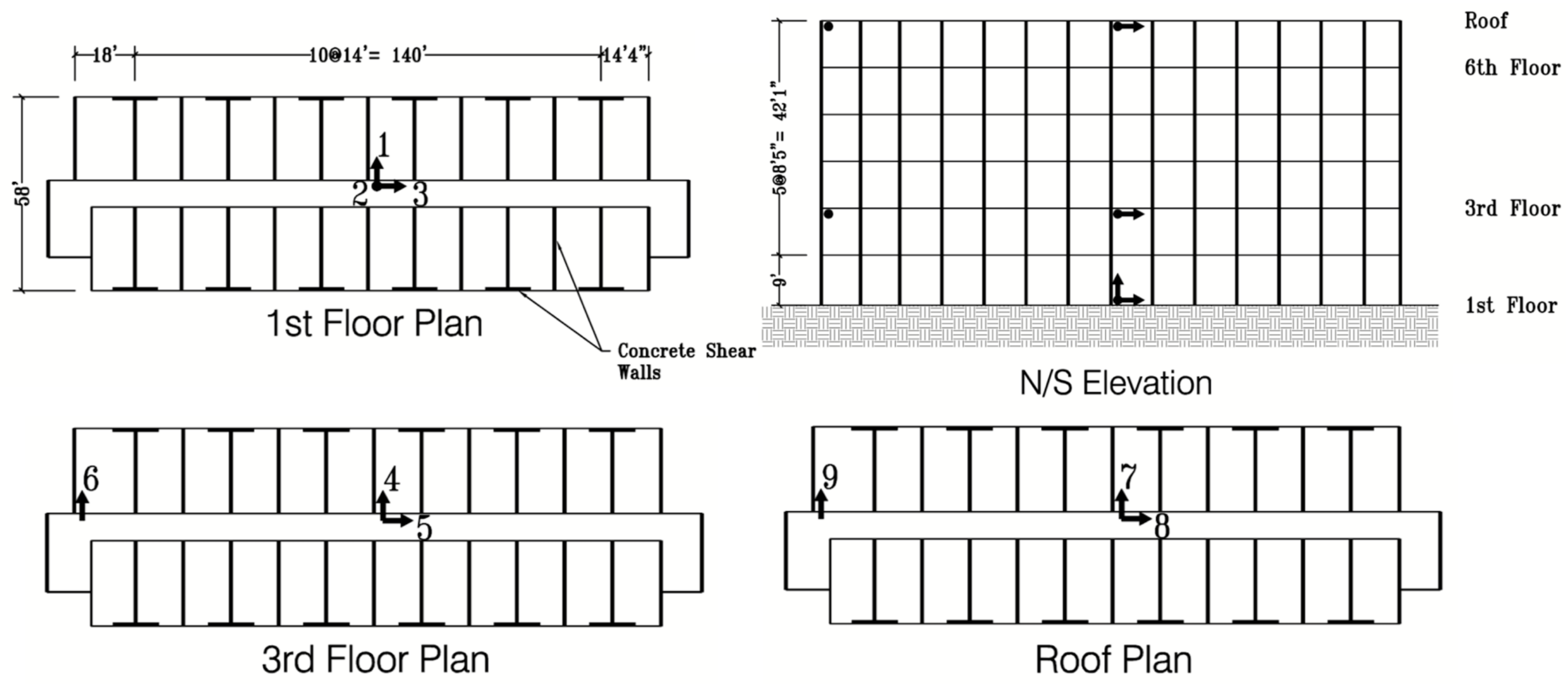}
		\caption{Sensor layout of the 6-story hotel in San Bernardino, California (Station Number: 23287) (\href{http://www.strongmotioncenter.org/}{http://www.strongmotioncenter.org/}).}
		\label{fig:hotel}
	\end{figure}

	\begin{table}[t!]
		\caption{Historical data information.} \vspace{6pt}
		\label{table1}
		\footnotesize
		\centering
		\begin{tabular}{cccccccccc}
			\hline
			Ind. & Earthquake                                                       & \begin{tabular}[c]{@{}c@{}}Epicenter\\ Distance\\ (km)\end{tabular} & \begin{tabular}[c]{@{}c@{}}PGA\\ (g)\end{tabular} & \multicolumn{1}{c|}{\begin{tabular}[c]{@{}c@{}}Peak Floor\\ Disp. (cm)\end{tabular}} & Ind. & Earthquake                                                      & \begin{tabular}[c]{@{}c@{}}Epicenter\\ Distance\\ (km)\end{tabular} & \begin{tabular}[c]{@{}c@{}}PGA\\ (g)\end{tabular} & \begin{tabular}[c]{@{}c@{}}Peak Floor\\ Disp. (cm)\end{tabular} \\ \hline
			\multicolumn{10}{c}{\cellcolor[HTML]{C0C0C0}Training Dataset}                                                                                                                                                                                                                                                                                                                                                                                                                                                                                                 \\ \hline
			1    & \begin{tabular}[c]{@{}c@{}}Borrego\\ Springs\\ 2010\end{tabular} & 102.5                                                               & 0.024                                             & \multicolumn{1}{c|}{0.406}                                                           & 7    & \begin{tabular}[c]{@{}c@{}}Beaumont\\ 2011\end{tabular}         & 22.6                                                                & 0.028                                             & 0.058                                                           \\
			2    & \begin{tabular}[c]{@{}c@{}}Devore\\ 2015\end{tabular}            & 18.6                                                                & 0.054                                             & \multicolumn{1}{c|}{0.319}                                                           & 8    & \begin{tabular}[c]{@{}c@{}}Lahabra\\ 2014\end{tabular}          & 60.7                                                                & 0.024                                             & 0.181                                                           \\
			3    & \begin{tabular}[c]{@{}c@{}}Fontana\\ 2014\end{tabular}           & 17.3                                                                & 0.034                                             & \multicolumn{1}{c|}{0.102}                                                           & 9    & \begin{tabular}[c]{@{}c@{}}Loma Linda\\ 2017\end{tabular}       & 4.8                                                                 & 0.025                                             & 0.047                                                           \\
			4    & \begin{tabular}[c]{@{}c@{}}Inglewood\\ 2009\end{tabular}         & 99.4                                                                & 0.008                                             & \multicolumn{1}{c|}{0.052}                                                           & 10   & \begin{tabular}[c]{@{}c@{}}Ontario\\ 2011\end{tabular}          & 27.8                                                                & 0.004                                             & 0.015                                                           \\
			5    & \begin{tabular}[c]{@{}c@{}}Ocotillo\\ 2010\end{tabular}          & 197.3                                                               & 0.007                                             & \multicolumn{1}{c|}{0.135}                                                           & 11   & \begin{tabular}[c]{@{}c@{}}Yorba Linda\\ 2012\end{tabular}      & 50.3                                                                & 0.003                                             & 0.021                                                           \\
			6    & \begin{tabular}[c]{@{}c@{}}San\\ Bernardino\\ 2009\end{tabular}  & 5.1                                                                 & 0.094                                             & \multicolumn{1}{c|}{0.852}                                                           &      &                                                                 &                                                                     &                                                   &                                                                 \\ \hline
			\multicolumn{10}{c}{\cellcolor[HTML]{C0C0C0}Validation Dataset}                                                                                                                                                                                                                                                                                                                                                                                                                                                                                               \\ \hline
			12   & \begin{tabular}[c]{@{}c@{}}Banning\\ 2016\end{tabular}           & 38.2                                                                & 0.019                                             & \multicolumn{1}{c|}{0.102}                                                           & 14   & \begin{tabular}[c]{@{}c@{}}Redlands\\ 2010\end{tabular}         & 11.5                                                                & 0.019                                             & 0.145                                                           \\
			13   & \begin{tabular}[c]{@{}c@{}}Banning\\ 2010\end{tabular}           & 38.9                                                                & 0.005                                             & \multicolumn{1}{c|}{0.017}                                                           & 15   & \begin{tabular}[c]{@{}c@{}}Trabuco\\ Canyon\\ 2018\end{tabular} & 40.9                                                                & 0.01                                              & 0.04                                                            \\ \hline
			\multicolumn{10}{c}{\cellcolor[HTML]{C0C0C0}Prediction Dataset}
			\\ \hline
			16   & \begin{tabular}[c]{@{}c@{}}Beaumont\\ 2010\end{tabular}          & 28.1                                                                & 0.005                                             & \multicolumn{1}{c|}{0.026}                                                           & 20   & \begin{tabular}[c]{@{}c@{}}Devore\\ 2012\end{tabular}           & 23.5                                                                & 0.01                                              & 0.052                                                           \\
			17   & \begin{tabular}[c]{@{}c@{}}Big Bear\\ Lake 2014\end{tabular}     & 33.4                                                                & 0.011                                             & \multicolumn{1}{c|}{0.107}                                                           & 21   & \begin{tabular}[c]{@{}c@{}}Loma Linda\\ 2013\end{tabular}       & 4.6                                                                 & 0.012                                             & 0.058                                                           \\
			18   & \begin{tabular}[c]{@{}c@{}}Fontana\\ 2015\end{tabular}           & 15.5                                                                & 0.01                                              & \multicolumn{1}{c|}{0.034}                                                           & 22   & \begin{tabular}[c]{@{}c@{}}Northridge\\ 1994\end{tabular}       & 117.4                                                               & 0.07                                              & 2.67                                                            \\
			19   & \begin{tabular}[c]{@{}c@{}}Loma Linda\\ 2016\end{tabular}        & 6.9                                                                 & 0.008                                             & \multicolumn{1}{c|}{0.034}                                                           & 23   & \begin{tabular}[c]{@{}c@{}}Landers\\ 1992\end{tabular}          & 79.9                                                                & 0.08                                              & 9.38                                                            \\ \hline
		\end{tabular}
	\end{table}

	\subsection{K-means clustering}
\begin{figure}[t!]
		\centering\includegraphics[width=1\linewidth]{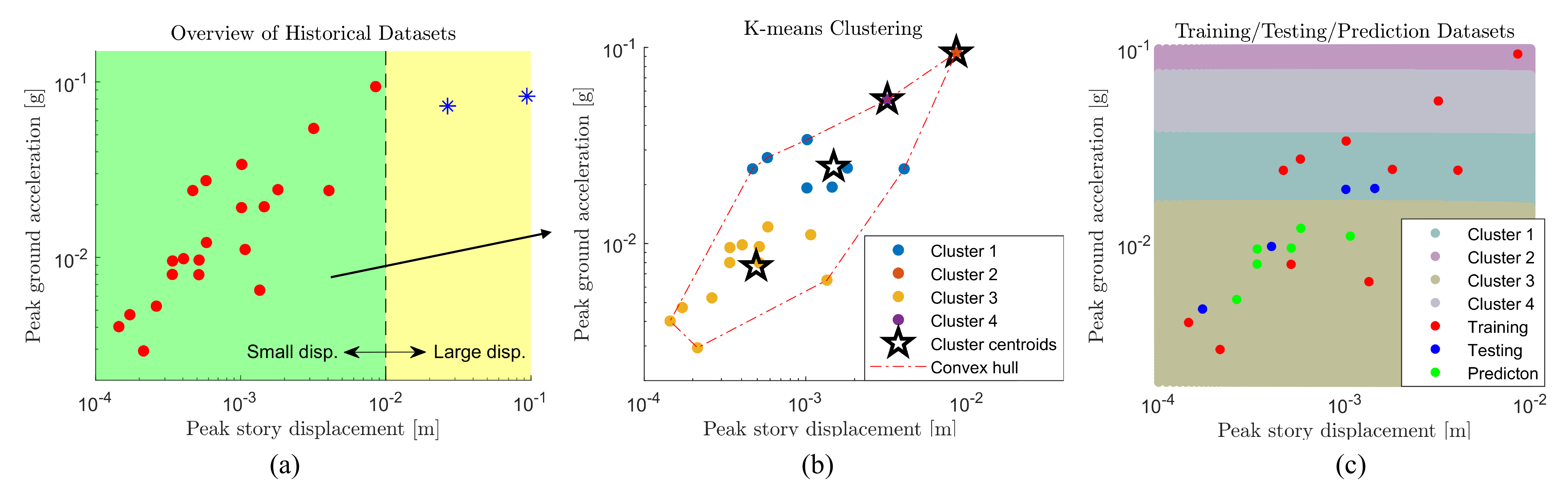}
		\caption{Data clustering using K-means clustering: (a) overview of historical datasets; (b) illustration of clusters (\(k = 4\)) and convex envelope; (c) training/validation/prediction datasets determined based on K-means algorithm and convex envelope.}
		\label{fig:k-means}
	\end{figure}

	Before clustering, the raw sensing measurements with different sampling rates and high-frequency noise were first preprocessed. The measured accelerations were passed through a 2-pole Butterworth high-pass filter with a cutoff frequency of 0.1 Hz to remove the low-frequency behavior. The displacement time-series are further obtained from the high-pass filtered accelerations and used to model the input-output (ground acceleration-structural displacements) relationship. The entire historical datasets summarized in Table \ref{table1} are divided into training, validation and prediction datasets using the K-means clustering and the convex envelope technique discussed in the following. Both training and validation datasets are considered as known where both input and output are fully given during the training process, while prediction dataset is considered as unknown where only ground acceleration is given. The historical data is clustered based on both input excitations and output structural responses. Figure \ref{fig:k-means}(a) shows the relationship of peak ground acceleration (PGA) versus the peak structural displacement for all datasets in logarithmic scale. It can be seen that the peak structural displacements of most samples are less than 1 cm as shown in the green region, which is considered as the boundary of interest for training. The other two samples in the yellow region which yield large displacement under Northridge and Landers earthquakes, are used to test the performance of the trained PhyCNN model under a larger level of response whose information might not be fully covered during training process. 
	
\begin{figure}[t!]
		\centering\includegraphics[width=0.65\linewidth]{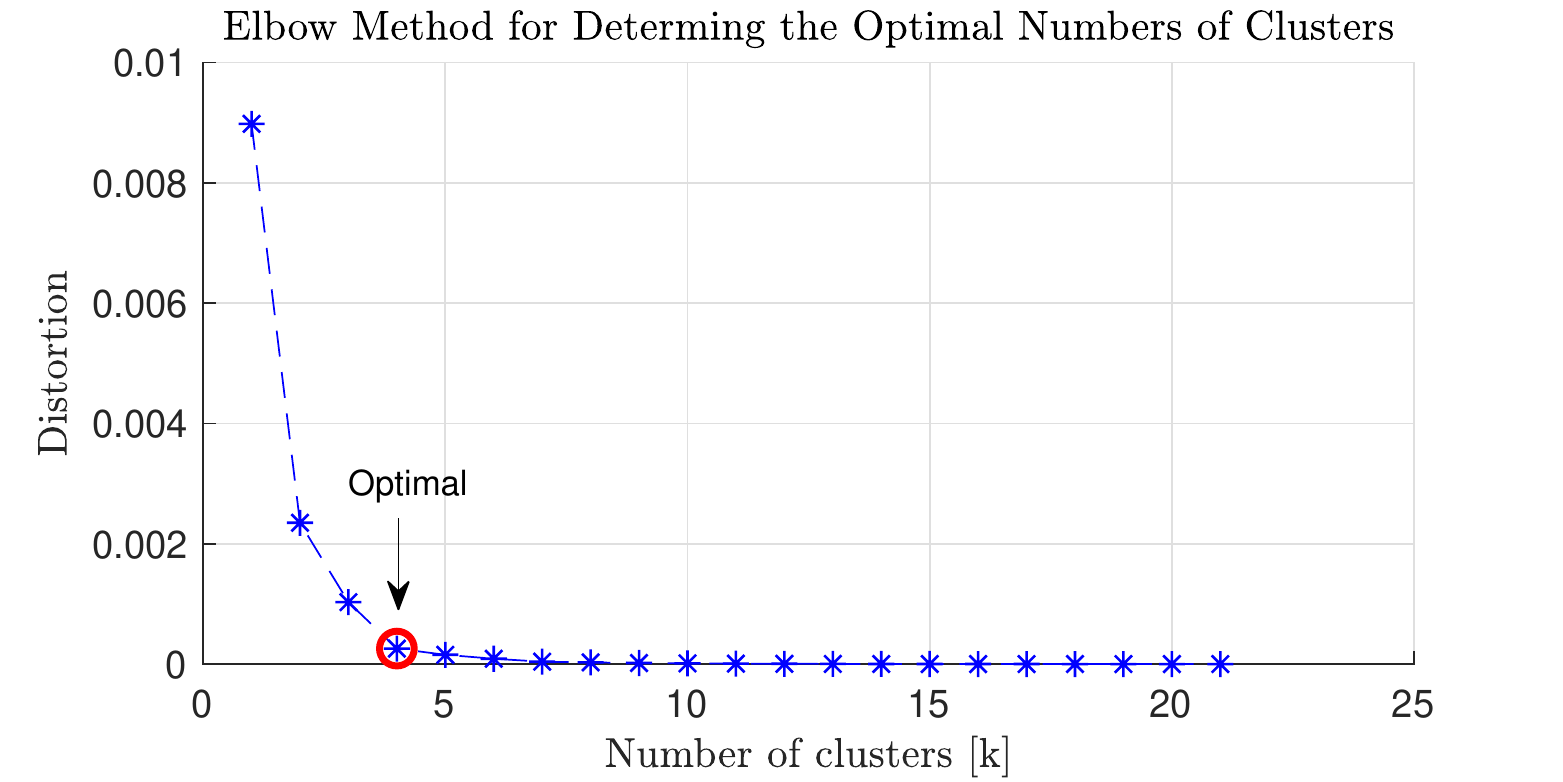}
		\caption{Identification of optimal number of clusters at the Elbow point.}
		\label{fig:elbow}
	\end{figure}

	The 21 samples within the boundary of interest (green region) are partitioned into several clusters using the K-means algorithm \cite{ng2012clustering} which is a popular data mining approach in the unsupervised learning setting that groups datasets into a certain number of clusters. It starts with the random selection of a set of k cluster centroids, e.g., $C = c_1, c_2, ..., c_k$. Next, each observation is assigned to the cluster, whose mean has the least squared Euclidean distance, given by 
	\begin{equation}\label{eq:kmean1}
	\operatorname*{arg\, min}_{c_i\in C} dist\left(c_i,x\right)^2
	\end{equation}
	where \(dist\) calculates the Euclidean distance. The new centroid is then determined by taking the mean of all the observations assigned to that cluster as shown in Eq. \eref{eq:kmean2}:
	\begin{equation}\label{eq:kmean2}
	c_i=\frac{1}{\left|S_i\right|}\sum_{x_i\in S_i} x_i
	\end{equation}
	where $S_i$ is the set of all observations assigned to the $i$th cluster. The algorithm converges when the cluster assignments no longer change. To determine the optimal number of clusters $k$ for the given observations, the elbow method is used which calculates the distortions for different numbers of clusters \cite{kodinariya2013review}. As shown in Figure \ref{fig:elbow}, the optimal value for $k$ is determined as $k$ = 4 where adding another cluster doesn't give much better modeling of data. The limitation of this method is that it cannot always be unambiguously identified. In such cases, other approaches such as Silhouette \cite{pollard2002method,kaufman2009finding} and Cross-validation \cite{smyth1996clustering} can be used to find the optimal numbers of clusters, which will be investigated in the future work.
	
	The 21 samples in the green region are divided into four clusters shown in Figure \ref{fig:k-means}(b). A total number of 11 datasets are selected for training by picking up the datasets on the convex envelope which defines the boundary of interest, as well as the datasets closest to the cluster centroids. The validation datasets can be determined by randomly and evenly picking from each cluster. In this study, since the datasets in Cluster 2 and Cluster 4 are insufficient, the validation datasets are selected from Cluster 1 and Cluster 3 (two for each). The rest 6 datasets plus the 2 datasets out of the boundary (in yellow region) are considered as the prediction datasets to demonstrate the performance of the proposed PhyCNN architecture both within and out of the boundary of interest. A summary of the datasets for training, validation and prediction purposes is illustrated in Figure \ref{fig:k-means}(c). The training and prediction performance for this 6-story hotel building is presented in the following subsection.
	
	\begin{figure}[t!]
		\centering
		\subfigure[]{\includegraphics[width=0.495\linewidth]{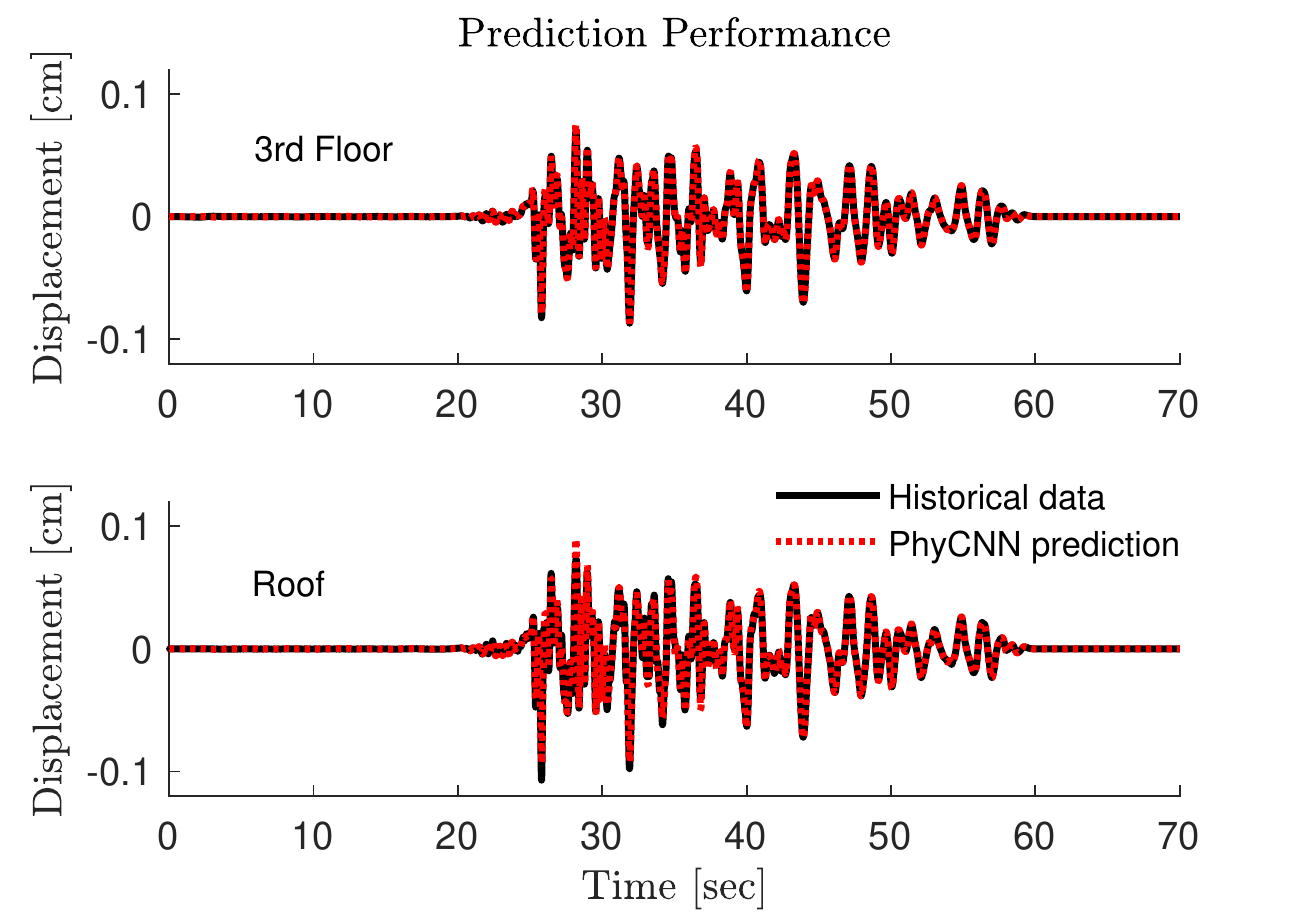}}
		\subfigure[]{\includegraphics[width=0.495\linewidth]{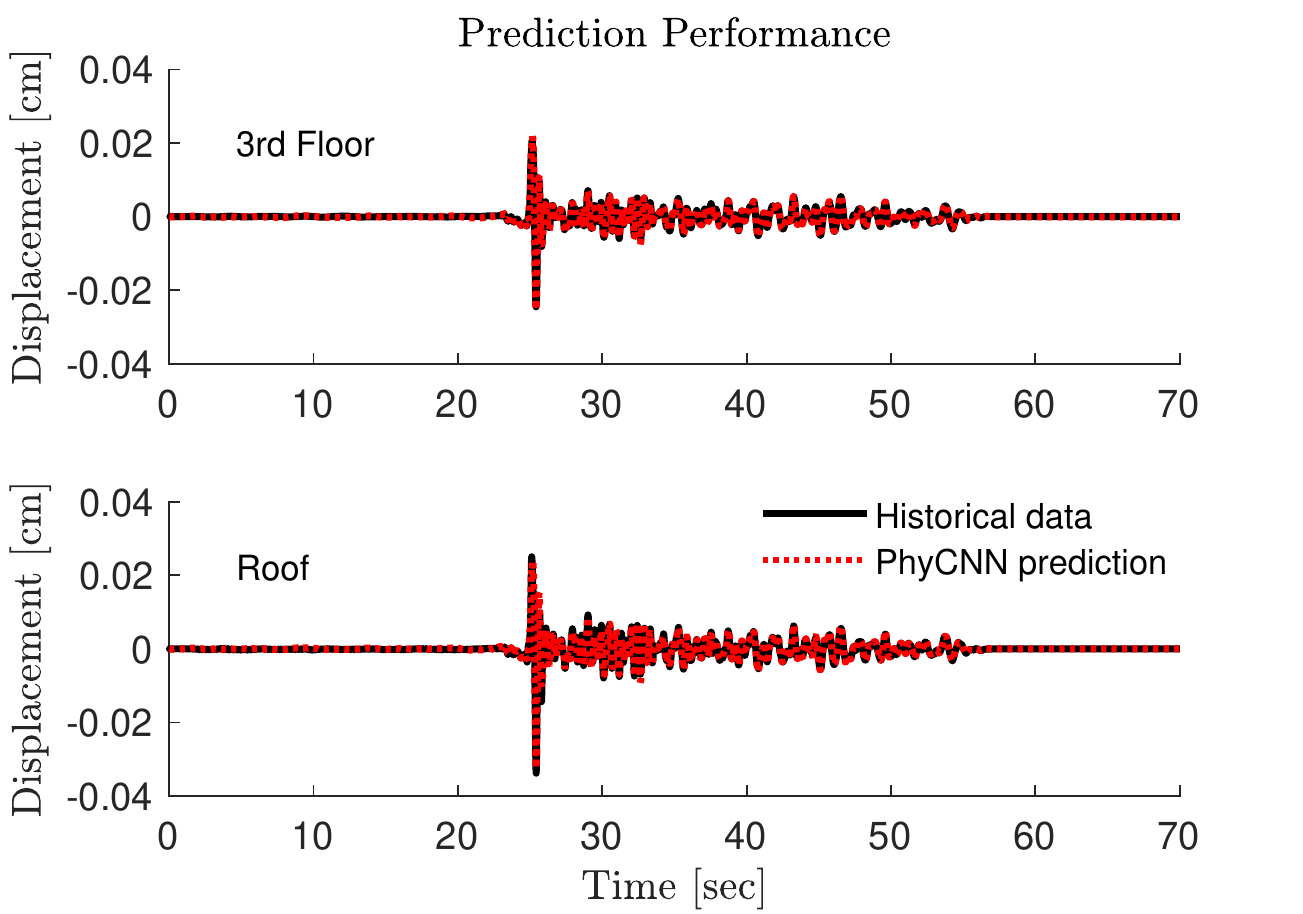}}
		\vspace{-18pt}
		\caption{Prediction performance of the proposed PhyCNN model.}
		\label{fig:sensing_pred}
	\end{figure}
	
	\subsection{Predicted displacements using PhyCNN}\label{4.3}
\begin{figure}[t!]
		\centering\includegraphics[width=0.8\linewidth]{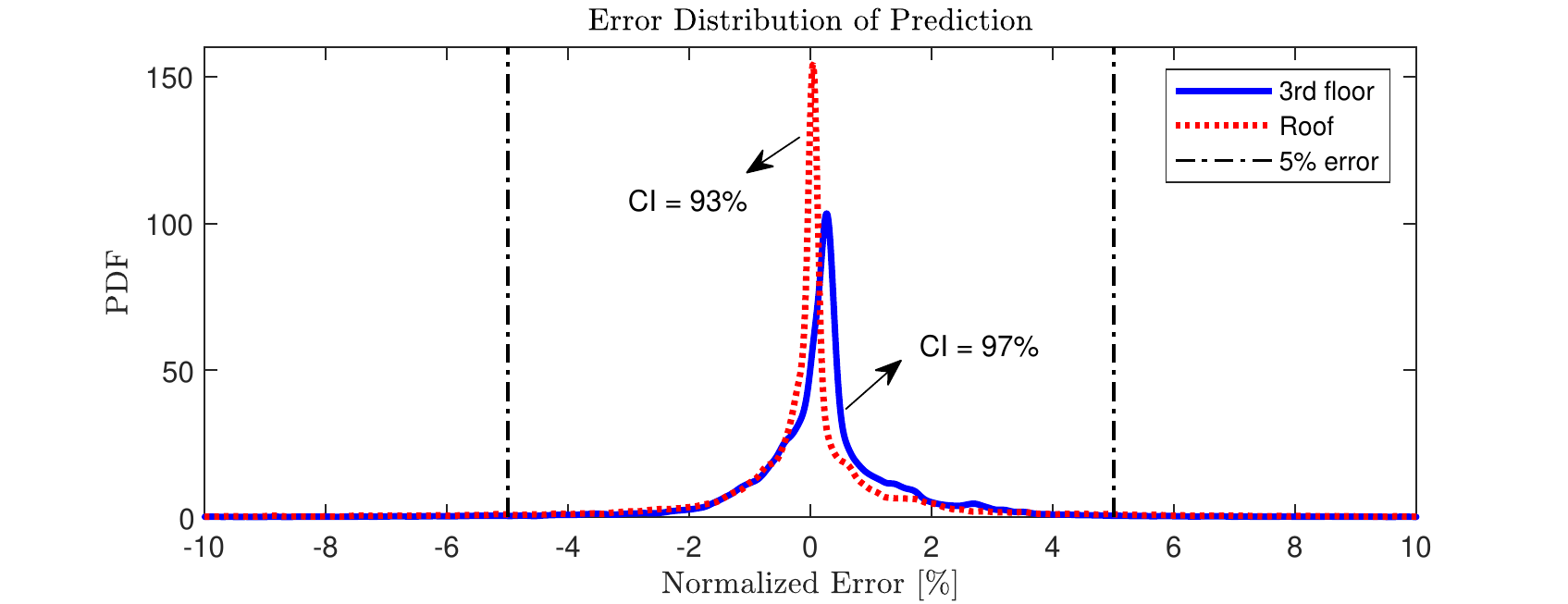}
		\caption{Error distribution of the prediction datasets using the proposed PhyCNN model.}
		\label{fig:pdf}
	\end{figure}

	The training and validation datasets discussed above are used to train the PhyCNN for the 6-story hotel building in San Bernardino, CA, consisting of 11 and 4 samples respectively, each of which contains sequences (with 7,200 data points) of ground motion accelerations as input and the story accelerations as the measurement data.
	During training, the training datasets are fed into the PhyCNN architecture used in the previous numerical example in Section \ref{sec3.2} with accelerations as the only field measurements. The trained PhyCNN model is then used to predict structural displacements under new earthquakes. By simply feeding a new ground acceleration into the trained PhyCNN model, it accurately predicts the structural displacements under that excitation. Figure \ref{fig:sensing_pred} shows the predicted story displacements of the 3rd floor and roof for Big Bear Lake 2014 and Loma Linda 2016 earthquakes. It can be clearly seen that the PhyCNN prediction matches the historical sensing data very well for earthquakes with different magnitudes and frequency contents. To better illustrate the prediction error, the probability density function (PDF) of the normalized error distribution defined in Eq. \eref{eq:pdf} is presented in Figure \ref{fig:pdf}. It can be seen that the prediction error is mainly located within 5\% for the 3rd floor and roof with a confidence interval (CI) of 97\% and 93\%, respectively. This demonstrates the high prediction accuracy of the proposed PhyCNN approach.
	\begin{equation}\label{eq:pdf}
	\mathcal{P}=\mathrm{PDF}\left\{\frac{y_{true\ }-y_{predict}}{\max{\left(\left|y_{true}\right|\right)}}\right\}\
	\end{equation}
	
\begin{figure}[t!]
		\centering
		\subfigure[]{\includegraphics[width=0.495\linewidth]{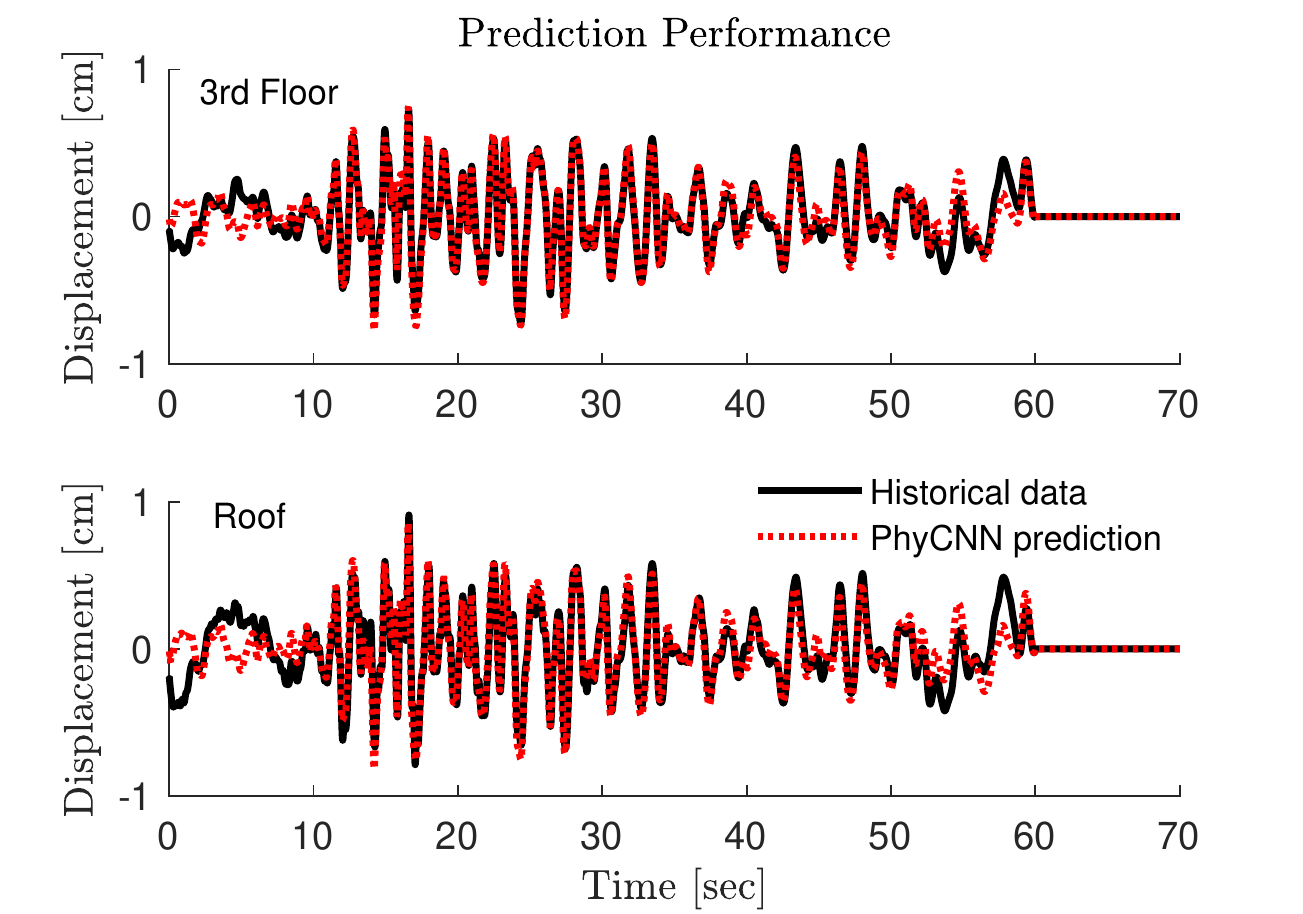}}
		\subfigure[]{\includegraphics[width=0.495\linewidth]{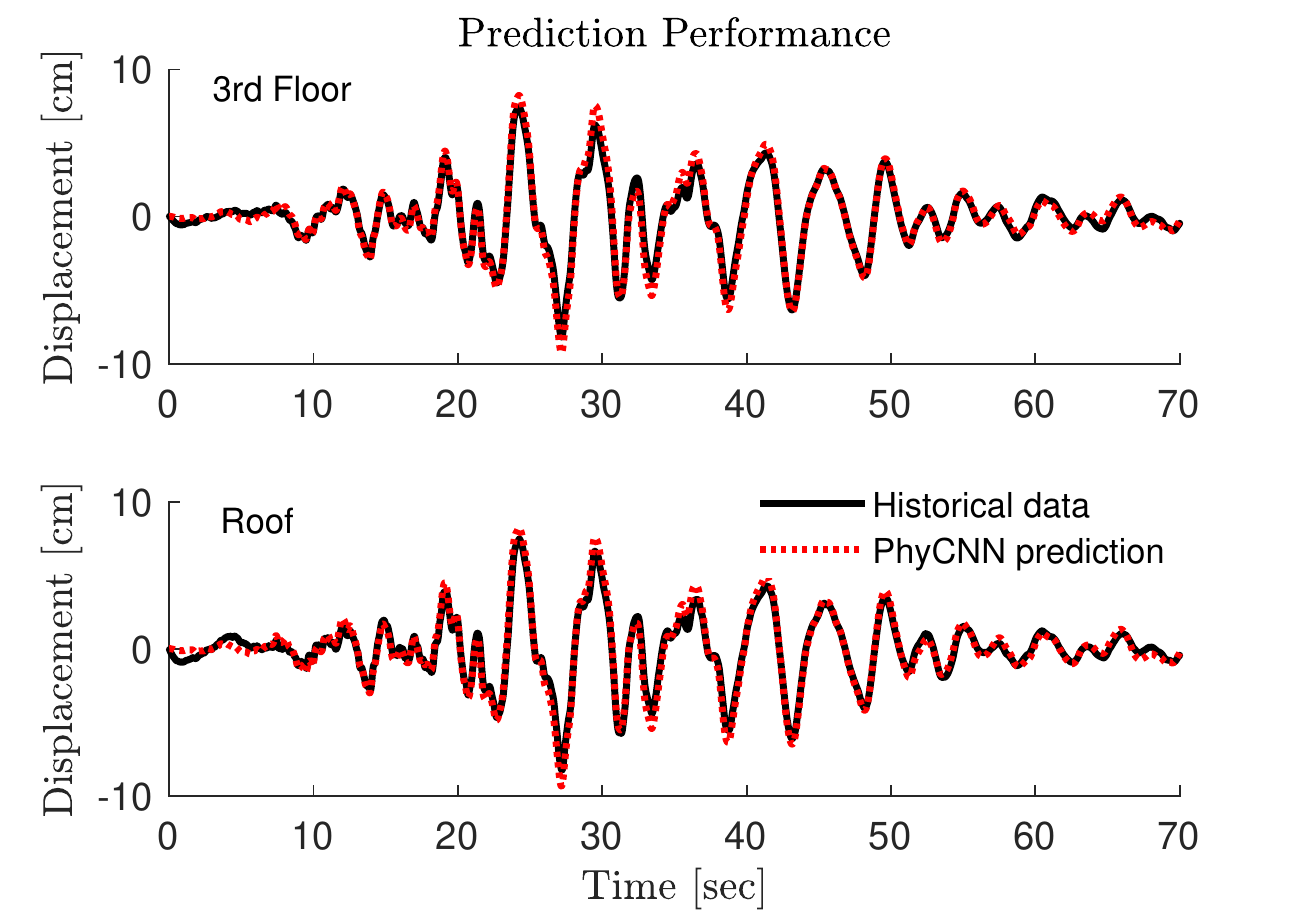}}
		\vspace{-18pt}
		\caption{Prediction performance of the proposed PhyCNN model under larger seismic intensities.}
		\label{fig:sensing_pred_intense}
	\end{figure}

	The extrapolation ability of the proposed PhyCNN is further verified using the two samples out of the boundary interest (in the yellow region as shown in Figure \ref{fig:k-means}(a)). Figure \ref{fig:sensing_pred_intense} shows the predicted structural displacements under Northridge 1994 and Landers 1992 earthquakes. It is observed that the proposed PhyCNN model is able to well predict structural responses for larger earthquakes, which offers confidence in applying the proposed method for building serviceability or fragility assessment.
	
	\subsection{Seismic Serviceability Analysis of the Building}
	The trained PhyCNN model is used as a surrogate model for structural seismic response prediction which can be further employed to develop fragility functions based on certain limit states for seismic serviceability analysis. The use of limit states in seismic risk assessment reflects the vulnerability of the building structures against earthquakes. The serviceability limit state, as one of the limit states, indicates the structural performance under operational service conditions and aims to minimize any future structural damage due to relatively low earthquakes \cite{dymiotis2004serviceability}. For serviceability assessment, the fragility function can be used to describe the probability of exceedance of the serviceability limit state for a specific earthquake intensity measure (IM). The probability of exceeding a given damage level (DL) is defined as a cumulative lognormal distribution function as follows
	\begin{equation}\label{eq:pdf_dl}
	P\left(\text{DL}\middle|\text{IM}=x\right)=\Phi\left[\frac{ln{\left(x/\mu\right)}}{\beta}\right]
	\end{equation}
	where $P\left(\text{DL}\middle|\text{IM}=x\right)$ denotes the probability that a ground motion with IM = $x$ exceeds a given performance level (e.g., serviceability limit state); $\Phi\left(\cdot\right)$ is the standard normal cumulative distribution function (CDF); $\mu$ is the median of the fragility function (the IM level with 50\% probability of exceeding the given DL); and $\beta$ is the standard deviation of the natural logarithm of the IM which describes the variability for structural damage states.
	
	To calibrate the fragility function, we need to estimate the parameters $\mu$ and $\beta$. Shinozuka et al. \cite{shinozuka2000nonlinear,shinozuka2000statistical} estimated $\mu$ and $\beta$ using the maximum likelihood estimation (MLE), denoted by $\mathcal{L}\left(\cdot\right)$. In the MLE approach, the damage state is related to a Bernoulli random variable. If the limit state is reached, $y_i$ is set as 1; otherwise, $y_i=0$. The likelihood function is given by 
	
	\begin{equation}\label{eq:mle}
	\mathcal{L}\left(\mu,\beta\right)=\prod_{i=1}^{N}{\Phi\left[\frac{\ln{\left(x_i/\mu\right)}}{\beta}\right]^{y_i}}\left[1-\Phi\left[\frac{\ln{\left(x_i/\mu\right)}}{\beta}\right]\right]^{1-y_i}
	\vspace{3mm}
	\end{equation}
	where $\Pi$ denotes a product over $N$ earthquake ground motions. Using an optimization algorithm, the two parameters $\mu$ and $\beta$ can be obtained when the likelihood function in the logarithmic space is maximized. 
	
	The serviceability assessment is conducted based on the performance-based engineering method. According to ASCE/SEI 41-06 standard \cite{asce2007seismic}, the building performance levels include operational, immediate occupancy, life safety, and collapse prevention. Both the operational and immediate occupancy performance level can be considered as the serviceability limit state. ASCE/SEI 41-06 also provides the recommended values for the maximum inter-story drift for each performance level and type of the structure. An alternative standard for fragility analysis is HAZUS \cite{fema2003multi}, in which the damage states of both structural and nonstructural components are defined. However, the inter-story drifts are typically unavailable due to the limitation of the sensor locations. Instead, the drift angle, defined as the ratio of the story deflection to story height, can be calculated and used as the threshold for serviceability assessment. Typical values of the drift angle for serviceability check lie in the range of [1/600, 1/100] for different building types and materials \cite{griffis1993serviceability}. In this paper, the threshold of the serviceability limit state is defined as 0.5\% for the maximum drift angle under the earthquake with a 10\% probability of exceedance in a 50-year period. For serviceability assessment of a building, the structural responses can be predicted under a group of new ground motions using the trained PhyCNN model. Thus, the fragility function is obtained using Eq. \eref{eq:pdf_dl} based on the serviceability limit state.
	
	\begin{figure}[t!]
		\centering
		\subfigure[]{\includegraphics[width=0.495\linewidth]{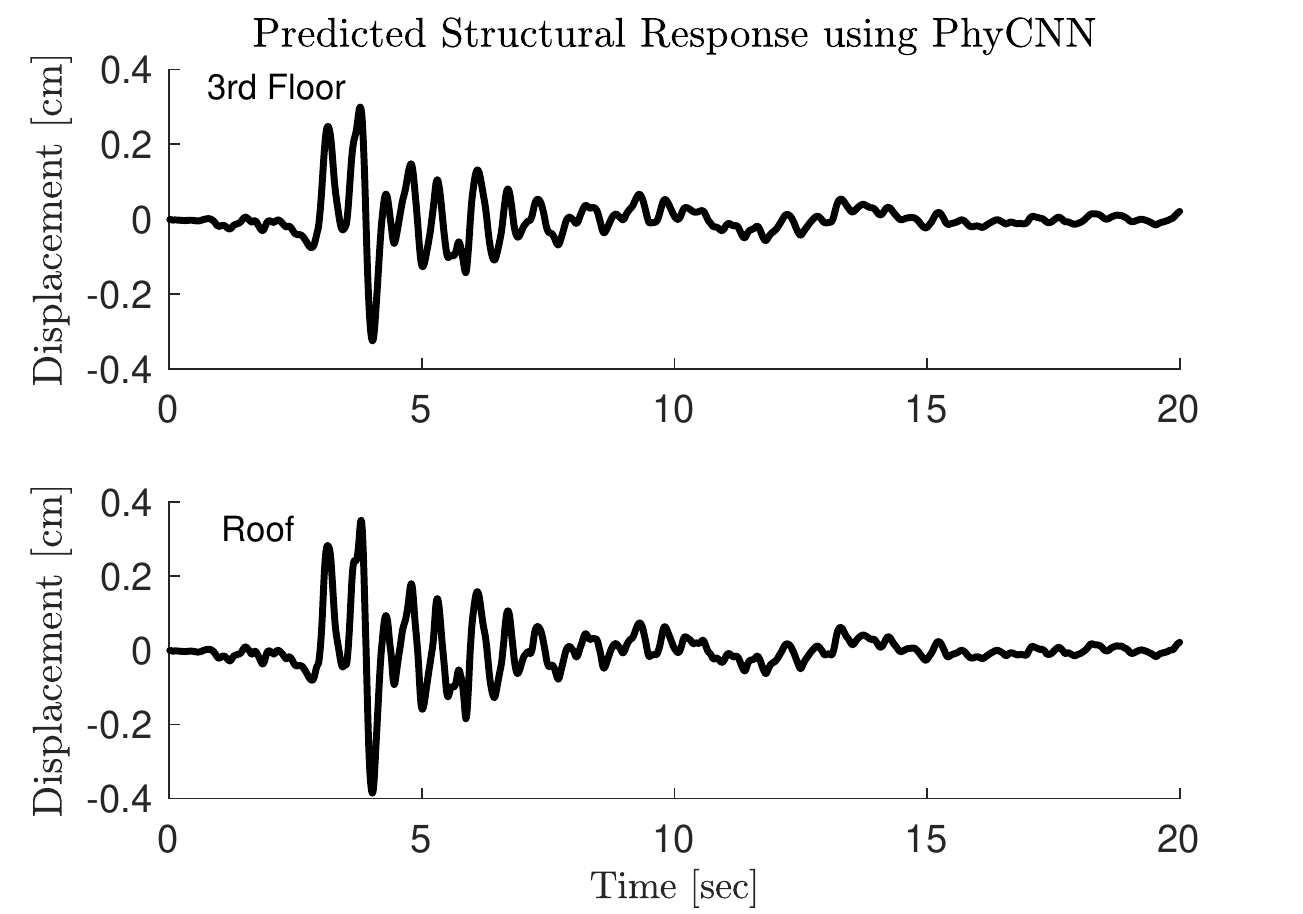}}
		\subfigure[]{\includegraphics[width=0.495\linewidth]{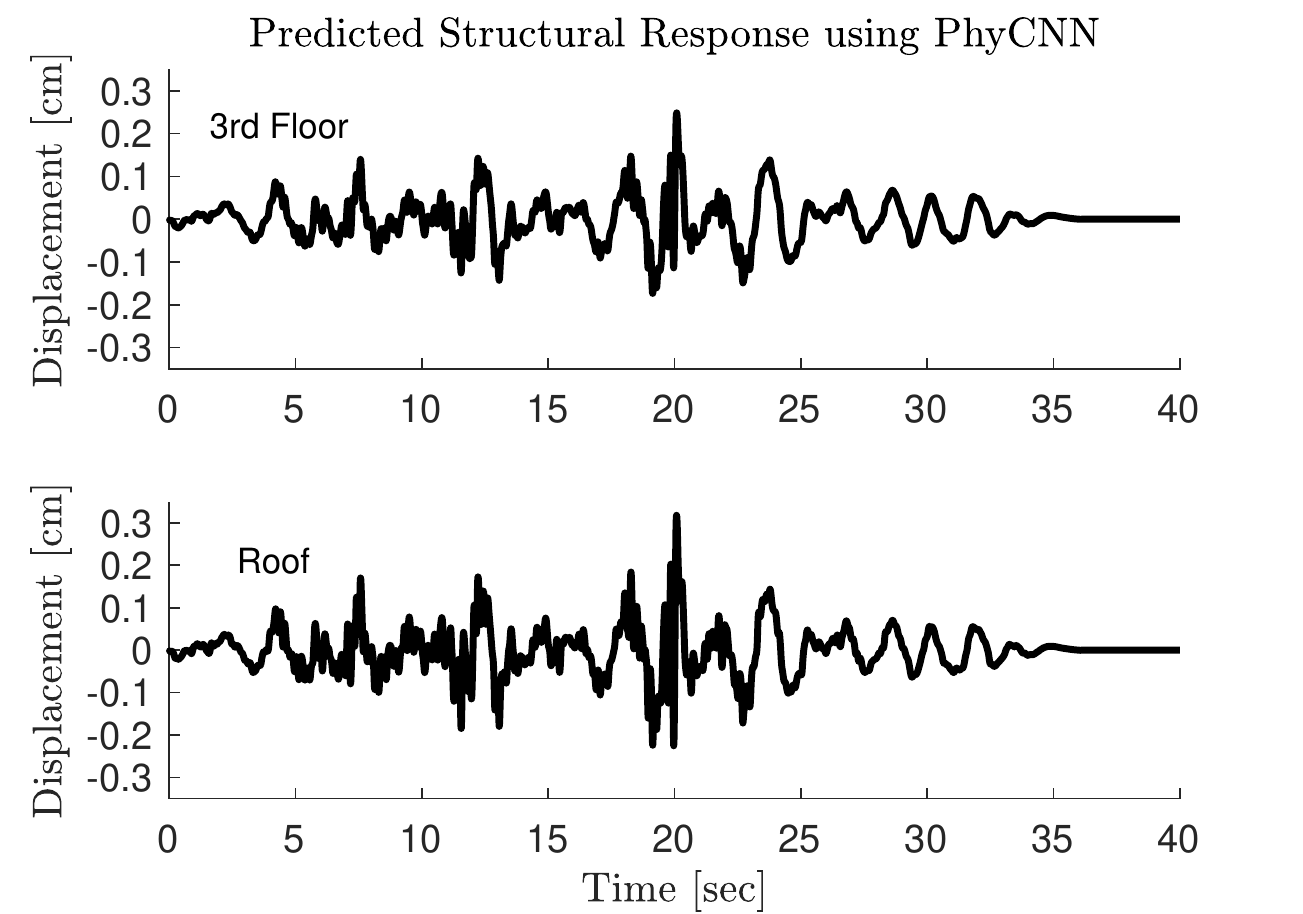}}
		\vspace{-18pt}
		\caption{Predicted structural responses under two example earthquakes using PhyCNN.}
		\label{fig:sensing_pred_new}
	\end{figure}
	
	\begin{figure}[t!]
		\centering\includegraphics[width=0.5\linewidth]{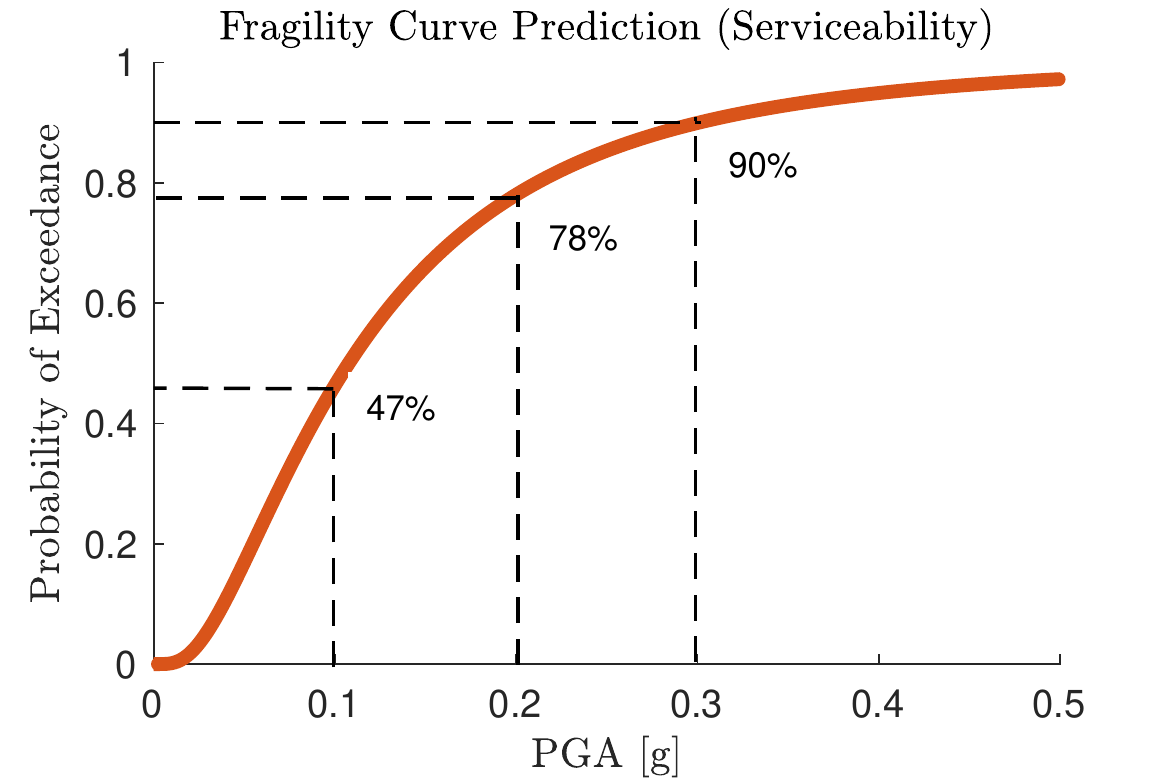}
		\vspace{-6pt}
		\caption{Predicted fragility curve of the serviceability limit state for the 6-story hotel building in San Bernardino using PhyCNN.}
		\label{fig:fragility}
	\end{figure}
	
	The serviceability of the 6-story hotel building in San Bernardino, CA is assessed based on the trained PhyCNN model described in Section \ref{4.3}. A suite of 100 ground motion records is input to the trained PhyCNN model to predict the structural displacements in an incremental dynamic analysis (IDA) setting. The 100 earthquake ground motions are selected from the PEER strong motion database \cite{chiou2008nga} in the area of San Bernardino with a 10\% probability of exceedance in 50 years. The mean response spectrum of the selected ground motion records matches the design spectrum of the 6-story hotel building. Figure \ref{fig:sensing_pred_new} shows the predicted displacements under two example new earthquakes. All the predicted displacements are then used to determine the fragility function with respect to the serviceability limit state. The fragility curve of the serviceability limit state is obtained based on Eq. \eref{eq:pdf_dl} and Eq. \eref{eq:mle} and shown in Figure \ref{fig:fragility}. It is seen that the probabilities of exceeding the serviceability limit state are around 47\%, 78\%, and 90\% for future earthquakes with PGA of 0.1g, 0.2g, and 0.3g, respectively. It is noted that the data-driven fragility curve can provide valuable information to guide the design of maintenance and rehabilitation strategies for the building. 
	
	\section{Conclusions}\label{sec_con}
	This paper presents a novel physics-guided convolutional neural network (PhyCNN) architecture to develop data-driven surrogate models for modeling/prediction of seismic response of building structures. The deep PhyCNN model includes several convolution layers and fully-connected layers to interpret the data, a graph-based tensor differentiator, and physics constraints. The key concept to leverage available physics (e.g., the law off dynamics) that can provide constraints to the network outputs, alleviate overfitting issues, reduce the need of big training datasets, and thus improve the robustness of the trained model for more reliable prediction. The performance of the proposed approach was illustrated by both numerical and experimental examples with limited datasets either from simulations or field sensing. The results show that the proposed deep PhyCNN model is an effective, reliable and computationally efficient approach for seismic structural response modeling. The trained model can further serve as a basis for developing fragility function for building serviceability assessment. Overall, the proposed algorithm is fundamental in nature which is scalable to other structures (e.g., bridges) under other types of hazard events.

	\section*{Acknowledgement}
	The authors would like to acknowledge the startup funds from the College of Engineering at Northeastern University, which support this study. The data and codes used in this paper will be publicly available on GitHub at \href{https://github.com/zhry10/PhyCNN}{https://github.com/zhry10/PhyCNN} after the paper is published.

	\bibliographystyle{elsarticle-num}
	\bibliography{refs}
	
\end{document}